%% file: paper.tex
\documentclass[12pt]{article}
\usepackage{graphicx}
\usepackage{epstopdf}
\usepackage{color}
\usepackage{cite}
\input lmacros

\makeatletter
\@addtoreset{equation}{section}

\makeatletter

\newcommand{\newsection}[1]{
\vspace{5mm}
\pagebreak[3]
\refstepcounter{section}
\setcounter{equation}{0}
\setcounter{subsection}{0}
\begin{flushleft}
{\large\bf \thesection. #1}
\end{flushleft}
\nopagebreak
\medskip
\nopagebreak}

\newcommand{\newsubsection}[1]{
\vspace{5mm}
\pagebreak[3]
\addtocounter{subsection}{1}
\addcontentsline{toc}{subsection}{\protect
\numberline{\arabic{section}.\arabic{subsection}}{#1}}
\noindent{\em 
\thesubsection. #1}
\nopagebreak
\vspace{2mm}
\nopagebreak}
\newcommand{\startappendix}{
\setcounter{section}{0}
\renewcommand{\thesection}{\Alph{section}}}

\newcommand{\Appendix}[1]{
\refstepcounter{section}
\vspace{10mm}
\pagebreak[3]
\setcounter{equation}{0}
\begin{flushleft}
{\large\bf Appendix \thesection: #1}
\end{flushleft}}

\def\baselinestretch{1.2}
\newcommand{\be}{\begin{equation}}
\newcommand{\ee}{\end{equation}}
\newcommand{\beq}{\begin{eqnarray}}
\newcommand{\eeq}{\end{eqnarray}}

\def\sec#1{Section \ref{#1}}
\def\fig#1{Fig \ref{#1}}
\def\req#1{(\ref{#1})}
\def\App#1{Appendix \ref{#1}}

\def\[{\left [}
\def\]{\right ]}
\def\({\left (}
\def\){\right )}
\def\ie{{\it i.e.}}
\def\etc{{\it etc.}}
\def\cf{{\it cf.}}

\def\a{\alpha}
\def\b{\beta}
\def\g{\gamma}

\def\e{\varepsilon}
\def\th{\theta}
\def\vp{\varphi}
\def\p{\partial}
\def\t{\tau}
\def\om{\omega}
\def\Om{\Omega}
\def\D{\Delta}

\def\CM{{\cal M}}
\def\CN{{\cal N}}
\def\CO{{\cal O}}

\def\R{{\bf R}}
\def\S{{\bf S}}
\def\A5S5{{\rm AdS}_5 \times \S^5}

\def\l{\ell}

\def\ket#1{\mid \! #1 \, \rangle}
\def\bra#1{\langle \, #1 \! \mid}
\def\iff{\Longleftrightarrow}

\def\ud{\dot{u}}
\def\vd{\dot{v}}
\def\xd#1{\dot{x}_{#1}}
\def\td{\dot{t}}
\def\rd{\dot{r}}
\def\zd{\dot{z}}
\def\ud{\dot{u}}
\def\vd{\dot{v}}
\def\Omd{\dot{\Om}}
\def\ed{\dot \e}
\def\dua{\( {\p \over \p u} \)^a}
\def\dva{\( {\p \over \p v} \)^a}
\def\dx#1{\( {\p \over \p x^{#1}} \)^a}
\def\dta{\( {\p \over \p t} \)^a}
\def\dya{\( {\p \over \p y} \)^a}

\def\bp{{\bf p}}
\def\bx{{\bf x}}
\def\Veff{V_{\rm eff}}

\def\nc{non-commutative}
\def\lnc{lightlike \nc}
\def\snc{spacelike \nc}

\title{{\bf Causal structures and holography }}

\author{Veronika E. Hubeny$^{\, a,b}$, Mukund Rangamani$^{\, a,b}$ and 
Simon F. Ross$^{\, a}$ 
\footnote{veronika.hubeny, mukund.rangamani, s.f.ross@durham.ac.uk}\\ \\
\small \sl $^a$Centre for Particle Theory \& Department of
Mathematical Sciences, 
\\[-1.5mm]
\small \sl Science Laboratories, South Road, Durham DH1 3LE, United Kingdom\\ 
\small\sl $^b$Department of Physics \& Theoretical Physics Group,
LBNL,\\[-1.5mm] 
\small\sl University of California, Berkeley, CA
94720, USA\\ }

\begin{document}
\setlength{\baselineskip}{16pt}
\begin{titlepage}
\maketitle
\begin{picture}(0,0)(0,0)
\put(335,340){hep-th/0504034}
\put(335,325){DCPT-05/15}
\put(335,310){UCB-PTH-05/07}
\put(335,295){LBNL-57390}
\end{picture}

\begin{abstract}
We explore the description of bulk causal structure in a dual field
theory. We observe that in the spacetime dual to a spacelike
non-commutative field theory, the causal structure in the boundary
directions is modified asymptotically. We
propose that this modification is described in the dual theory by a
modification of the micro-causal light cone. Previous studies of this
micro-causal light cone for spacelike non-commutativite field theories
agree with the expectations from the bulk spacetime. We describe the
spacetime dual to field theories with lightlike non-commutativity, and
show that they generically have a drastic modification of the light
cone in the bulk: the spacetime is non-distinguishing. This means
that the spacetime while being devoid of closed timelike or null curves,  
has causal curves that are ``almost closed''. We go on to show that
the micro-causal light cone in the field theory agrees with this
prediction from the bulk.
\end{abstract}
\thispagestyle{empty}
\setcounter{page}{0}
\end{titlepage}

\renewcommand{\baselinestretch}{1.2}  
\newsection{Introduction}\label{intro}

The advent of the AdS/CFT correspondence \cite{Aharony:1999ti} has
revolutionised our understanding of quantum gravity, and has led to
important insights into both gravity and strongly coupled gauge
dynamics. However, a 
considerable number of important conceptual questions remain
open, especially those pertaining to a detailed understanding of gravitational 
dynamics. In particular, we still lack full understanding of how 
field theory on
a fixed background can be approximated by a local theory
with dynamical metric in the bulk.

One major issue is that the bulk spacetime has some dynamically
determined causal structure, whereas the dual field theory lives in flat
space, with the usual fixed light cone. The description of the bulk
causal structure from the dual point of view has been the subject of a
number of investigations in the AdS/CFT correspondence. Bulk
causality in global AdS spacetime is consistent with boundary
causality~\cite{Horowitz:1999gf}; for example, propagation through the
bulk is never faster than propagation along the boundary (this was
extended to asymptotically AdS spacetimes satisfying the weak energy
condition in~\cite{Gao:2000ga}). When we restrict to Poincar\'e-invariant
states, the light cone in the boundary directions will automatically
agree in the bulk and boundary, and the focus is on understanding the
causal structure in the bulk radial direction. In~\cite{Kabat:1999yq},
Kabat and Lifschytz proposed a scale/radius description, where
causality in the radial direction is enforced by a speed limit on
varying scale size in the field theory. If we violate Poincar\'e
invariance, for example by considering a black hole spacetime in the
bulk (corresponding to finite temperature in the field theory), the
bulk causal structure will not in general agree with the field theory
one even in the boundary directions. The extension of Kabat \&
Lifschytz's analysis to such cases was considered
in~\cite{Gregory:2000an}. However, for asymptotically AdS spacetimes,
the causal structure changes only in the interior of the spacetime,
which makes it difficult to pose sharp questions about the
interpretation of these changes in the field theory.

In this paper, we will consider extreme examples of changes in the
bulk causal structure: spacetimes dual to non-commutative field
theories. Since the non-commutativity changes the structure of the
field theory even in the ultraviolet, the bulk spacetime is no longer
asymptotically AdS. The point of interest to us is that the causal
structure of the bulk spacetime is drastically modified: in the limit
as we approach the boundary, the light cone in the boundary directions
ceases to depend on the directions in which non-commutativity is
turned on. Thus, propagation through the bulk can take us outside the
flat space light cone in these directions. We want to understand how
this is described from the field theory point of view. One would
expect that the non-locality of the non-commutative field theory plays
a crucial role, and we will argue that this is indeed the
case. In~\cite{Chaichian:2002vw,Alvarez-Gaume:2003mb,Chu:2005nb}, 
it was argued that
the non-locality leads to modifications of the micro-causal light
cone: in particular, it does not agree with the naive causal structure
of the background flat space the field theory is defined in. We will
show  in a simple toy example, scalar
non-commutative field theory, that these modifications 
agree with the light cone in the
boundary directions of the bulk spacetime. Extending this result to 
\nc\ gauge theories is somewhat tricky owing to issues related to 
absence of local gauge invariant observables. Nevertheless, we will 
argue based on the known properties of correlation functions of 
gauge invariant operators, that we 
indeed expect to see a modified causal structure in the field theory.

We study the modifications of the causal structure for both the
spacelike non-commutativity, previously studied
in~\cite{Alvarez-Gaume:2003mb,Chu:2005nb}, and for lightlike
non-commutativity. The lightlike case is especially interesting
because the corresponding spacetime geometry has a very radical
modification of its causal structure. The spacetime becomes {\it
non-distinguishing}, meaning that distinct points of the spacetime
have the same causal future and past. Such a spacetime `almost has'
closed timelike curves (CTCs); it is in the borderline area between
being clearly causally well-behaved and clearly pathological. Since
the discovery of supersymmetric spacetimes with
CTCs~\cite{Gauntlett:2002nw}, there has been a lot of discussion of
whether spacetimes with CTCs are admissible backgrounds in string
theory. Most of the discussion has centered on some string version of
chronology protection; for example~\cite{Gibbons:1999uv} related
chronology violation in AdS/CFT to unitarity loss,
while~\cite{Boyda:2002ba} argue for a holographic chronology
protection mechanism. In light of the latter proposal that holography
could `cut off' the region of CTCs, it is remarkable to observe that
lightlike non-commutativity provides an example where a
non-distinguishing spacetime has a well-behaved holographic dual
description. We will suggest that the non-distinguishing character of
this spacetime is reflected in the micro-causal structure of the
lightlike non-commutative field theory. Understanding this case may
also be helpful for understanding the dual description of plane wave
spacetimes, as they are also on the borderline (although they are
distinguishing, they are not globally hyperbolic, and do not have
Cauchy surfaces). Non-distinguishing examples of pp-waves have also
recently been
constructed~\cite{Flores:2002fx,Hubeny:2003sj,Flores:2004dr}.

In the next section, we will review the construction of the dual
spacetime for a \snc\ deformation of $\CN =4$
SYM~\cite{Hashimoto:1999ut,Maldacena:1999mh}, and discuss the features
of the bulk light cone. In \sec{bdysp}, we consider the
micro-causality in spacelike non-commutative field theory. We review
the argument of~\cite{Alvarez-Gaume:2003mb,Chu:2005nb} that the
micro-causality condition for \snc\ field theories is to be imposed
inside a `light wedge'. In \cite{Alvarez-Gaume:2003mb, Chaichian:2004hb} 
it was argued that for a \nc\ field theory on $\R^{3,1}$ with $\th^{23} \neq 0$, the
breaking of Lorentz symmetry by non-commutativity would enlarge the
region where the commutator of fields is non-vanishing from the usual
light cone to a light wedge respecting the unbroken $SO(1,1) \times
SO(2)$ symmetry. As discussed in~\cite{Chu:2005nb}, the non-local
character of the field theory plays an essential role in this
enlargement. We show that this light wedge agrees with the bulk
results and suggest a framework for exploring these issues 
in \nc\ gauge theories.

We then turn to the consideration of lightlike non-commutativity. In
\sec{nmt} we construct the bulk spacetime by applying the ``Null
Melvin Twist'', a solution-generating technique discussed in
\cite{Alishahiha:2003ru, Gimon:2003xk}. We get our solution by
applying this twist to the extremal D3-brane, and then taking a
decoupling limit.  We demonstrate that the spacetime is
non-distinguishing using the arguments
of~\cite{Flores:2002fx,Hubeny:2003sj}, and briefly sketched in the appendix.
  We then show in \sec{ll} that
this spacetime is the holographic dual description of a theory with
lightlike non-commutativity. (This solution and its dual description
were previously obtained in~\cite{Alishahiha:2000pu}.) It was shown
in~\cite{Aharony:2000gz} that theories with lightlike
non-commutativity are well-defined quantum field theories, albeit with
non-local interactions. We thus have an example where a
non-distinguishing spacetime has a non-perturbative quantum
description.  We go on to show that any spacetime dual to a generic
theory with lightlike non-commutativity will be non-distinguishing.
Extending our arguments from the spacelike case, we are able to
demonstrate that the micro-causal structure in a simple scalar field
theory with lightlike non-commutativity reproduces the predicted
causal structure inferred from these non-distinguishing spacetimes.
Commutators of local operators can be shown to vanish only at equal
light cone time, respecting the unbroken Galilean subgroup of the
Lorentz group (in the presence of lightlike non-commutativity).  
We further argue that subtleties relating to gauge invariance 
will not change the result significantly in the case of \lnc\ gauge 
theories and conclude with a brief discussion. Some details of the field theory
calculations and more general solutions obtained by the null Melvin
twist are described in the appendices. A short summary of the essential 
physical ideas can be found in~\cite{Hubeny:2005ab}.

\newsection{Spacelike non-commutativity: bulk light cone}
\label{splw}

We begin by discussing the situation in the case of \snc\ field
theories. In this section, we explore the causal properties of the 
dual spacetime, and note that it has a deformed light cone even in the 
asymptotic region. We consider geodesics which remain at large radius to 
study the features of this light cone.

The string frame metric for the spacetime dual to \snc\ $\CN=4$ SYM is
\cite{Hashimoto:1999ut, Maldacena:1999mh}
\eqn{spncmet}{
ds^2 = {r^2 \over R^2} \, \(-dt^2 + dx_1^2 \)  + 
{R^2 \, r^2 \over R^4 + \g^2 \, r^4} \, \( dx_2^2  + dx_3^2 \) + 
{R^2\over r^2}\, dr^2 + R^2 \, d\Omega_5^2 \ . } 
There are non-trivial 3-form and 5-form fluxes in the above background
and a varying dilaton, but these are not going to be relevant to our
present discussion. $R$ as usual denotes the AdS radius and $\g$ is
related to the \nc\ parameter in the field theory as $\th^{23} = \g$. 
 
This dual geometry is obtained by thinking of the \snc\ Yang-Mills
theory as arising as the low-energy limit of the theory on D3-branes
in a constant background B-field. The supergravity configuration
sourced by these D3-branes can be obtained by introducing the B-field
through a twisting operation. To twist, start from the D3-branes in empty
space, T-dualize to D2-branes, and then consider instead the
compactification of the D2-brane solution on a tilted
torus. T-dualizing along this tilted direction will give D3-branes in
a B-field background. Applying this twisting operation to the D3-brane
metric, and then taking the decoupling limit to focus on the
near-horizon geometry, gives the spacetime \spncmet.

While the geometry \spncmet\ is close to $\A5S5$ for small $r$, it has
vastly different asymptotics. In particular, $g_{x_i x_i} \to 0 $ as
$r \to \infty$ for $i=2,3$, which will give rise to the deformation of
the light cone we are interested in. This makes it hard to define a
boundary for the spacetime, but we will show later that the
deformation of the light cone is nonetheless reflected in field theory
micro-causality behaviour.

Suppose we look at the metric induced on a fixed $r= r_0$ surface 
with $r_0 \gg R$, \ie, we think of the bulk defining an UV regulated 
version of the \nc\ field theory. In this case, the induced metric on the 
surface (ignoring the $\S^5$ part) reads:
\eqn{indmet}{
ds^2 \approx {r_0^2 \over R^2} \, \(-dt^2 + dx_1^2 + {R^4 \over
\g^2 \, r_0^4} \, \(dx_2^2 + dx_3^2 \) \) \  .}
At any fixed value of $r_0$, we can rescale the coordinates $(t,x_i)$
to convert \indmet\ into a flat metric on $\R^{3,1}$. However, the
scaling is non-homogeneous, and in particular, as we remove the cutoff
we will have to scale the shrinking directions $(x_2,x_3)$ by a
diverging amount, whereas we would expect the field theory background
to be defined by a homogeneous rescaling of coordinates, as in the
usual commutative case. This would lead us to conclude 
that the relevant light cone for the field theory is  independent of 
the non-commuting directions $\{x_2,x_3\}$.
Alternately, as is apparent from the full metric \spncmet, the 
metric factor in front of the non-commuting directions falls off faster 
than that in front of the $\S^5$ part, and therefore should not 
contribute to the light cone asymptotically.

We would like to argue that this change in the bulk light cone is reflected
in observable quantities in the dual field theory. One approach would
be to study the two-point function in the boundary, which should be
determined, by the usual logic, by the bulk-to-bulk propagator for the
appropriate supergravity field in the background \spncmet\ in the limit 
$r \to \infty$. By examining the behaviour of the bulk Greens functions 
we should be able to extract the detailed properties of the bulk light cone, 
and in particular its asymptotic behaviour. However,
the Klein-Gordon equation is rather formidable (it 
is related to the Mathieu equation \cite{Maldacena:1999mh}), and so 
explicit determination of the propagator is difficult. Nevertheless,  
we can approximate it by studying the geodesics in the 
bulk spacetime. We will now show that the bulk light cone 
indeed degenerates into a light wedge asymptotically  in the 
geodesic approximation.

Ignoring motion on the $\S^5$ part of the geometry and setting $R =1$ 
for simplicity, we have the equation for timelike geodesics
\eqn{timenc}{
-1 = r^2 \, \( - \td^2 + \xd1^2 \) + {r^2 \over 1 + \g^2 \, r^4} \, \(
\xd2^2 +\xd3^2 \) 
+ {\rd^2 \over r^2} \ , }
where $ ^{.} \equiv {d \over d \t}$, with $\t$ the affine parameter
along the geodesics.  The Killing symmetries $\dta$ and $\dx{i}$
determine the conserved energy and momenta $(p_t, p_i)$,
respectively. The geodesic motion reduces to a classical problem for
the zero energy trajectories of a particle in an effective potential
$\Veff(r)$:
\eqn{geodt}{\eqalign{
 \rd^2 + \Veff (r) &= 0, \cr
\td  = {p_t \over r^2} \ , \qquad & \xd1 = {p_1 \over r^2} \ , \qquad 
\xd{i} = {1 +\g^2 \, r^4 \over r^2} \, p_i , \cr
\Veff(r)   = r^2 +&\[- p_t^2 +p_1^2 + \(1+ \g^2 \, r^4 \) \, \(p_2^2
 + p_3^2 \) \] 
.}}

Let us consider geodesics that travel an infinitesimal distance in the
radial direction.  Choose $ r = r_0 \, \( 1 -\e \)$ with $\e \ll 1$
such that $\Veff(r_0) =0$. Then we have
\eqn{infgds}{
r_0^2 \, \ed^2  \approx 2\, r_0^2\,  \e \( 1 + 2 \, \g^2 \, r_0^2 \, \(p_2^2 +
p_3^2 \) \)  
\equiv 2 \, \e \, \a .} 
This allows us to approximately integrate the geodesic equations to
obtain, for instance,
\eqn{intdt}{
\D t = \int \, d \t \, {p_t \over r^2} \approx  {p_t \over r_0^2} \, \int d\t
={p_t \over r_0 \sqrt{\a}}\, \sqrt{2\, \e} }
and
\eqn{intxdt}{
\D x_2 \approx {p_2 \over r_0 \, \sqrt{\a}} \, \(1 + \g^2 \,r_0^4 \)\, 
\sqrt{2\, \e} .
}
One can clearly make the distance travelled by the geodesics in
individual directions large, by choosing appropriate values for the
momenta. However, the motion is confined to the region in the
immediate vicinity of the classical bulk light cone, as the proper
interval remains small:
\eqn{indis}{\eqalign{
-\D t^2 + \D x_1^2 + { \D x_2^2 + \D x_3^2 \over 1 + \g^2 \, r_0^4} & 
\approx
{2\, \e \over \a \, r_0^2} (-p_t^2  + p_1^2 + (1+\g^2 \, r_0^4) (p_2^2 +
p_3^2) )\cr 
& \approx -{2\, \e \over \a}  \ , 
}}
where we made use of the fact that $\Veff(r_0) = 0 $.

Thus, bulk geodesics can relate points inside the light cone given by
\eqn{blwedge}{
-\D t^2 + \D x_1^2 + { \D x_2^2 + \D x_3^2 \over 1 + \g^2 \, r_0^4} = 0,
}
and as we remove the cutoff by taking $r_0 \to \infty$, the light cone
in the boundary directions will approach a `light wedge': 
\eqn{blwedgestr}{
 -\D t^2 + \D x_1^2 =0 \ .
 }
This is the advertised modification of the bulk causal structure. 
Although we have restricted our attention to bulk geodesics, 
we will have essentially obtained the same answer by looking
at say the free scalar propagator for \spncmet. Having obtained 
a prediction from the bulk perspective, we now turn to
looking for signs of this change in the causal structure in 
the field theory.

\newsection{Micro-causality in \nc\ field theories}
\label{bdysp}

The field theory dual to the geometry \spncmet\ is  $\CN =4$ SYM theory 
in flat four-dimensional spacetime, with a constant non-commutativity
parameter $\th^{23} = \g$. We want to show that the above `light wedge'
structure is also naturally reflected in this field theory. As we are
dealing with a Lorentz non-invariant non-local field theory, it is not
{\it a priori} obvious what the causal relations in the field theory
are. We will see that if we define the light cone of the field theory
to be the micro-causal light cone, that is, the boundary of the region
where the commutator of elementary fields is non-vanishing, then it
will agree with the light wedge predicted from the bulk. 

\newsubsection{Non-commutative scalar field dynamics: a toy model}

The micro-causal light cone for \snc\ field theories was studied in
\cite{Alvarez-Gaume:2003mb}, and further clarified recently in
\cite{Chu:2005nb}. In \cite{Alvarez-Gaume:2003mb}, a heuristic
argument was given for a modification of the micro-causality
condition. The non-commutativity in a spatial $\R^2 \subset \R^{3,1}$,
$\th^{23} \neq 0$, breaks Lorentz invariance in the field theory from
$SO(3,1) \to SO(1,1) \times SO(2)$.  As a result it was suggested that
one should not demand causal behaviour with respect to the full
$SO(3,1)$ invariance, but rather only with respect to the smaller symmetry
$SO(1,1)$. This corresponds to the invariance of the light wedge
obtained above (although the reduced symmetry is not sufficient to fix
the form of the light cone). 

That is, the authors of \cite{Alvarez-Gaume:2003mb} propose that 
instead of requiring that the fields (labeled collectively as $\Phi(x)$) 
commute (or anti-commute for fermions) across spacelike separated 
points,
\eqn{umicro}{
\langle \, [\Phi(x), \Phi(y)]_{\pm} \, \rangle 
= 0 \, \qquad {\rm for } \;\; (x -y)^2 >  0 \ . }
the appropriate micro-causality condition for \nc\ field theories
would be to only impose
\eqn{spmicro}{
\langle \, [\Phi(x), \Phi(y)]_{\pm} \, \rangle 
= 0 \, \qquad { \rm for } \;\; (x -y)_c^2 >  0 \ . }
where $(x -y)_c^2$ is the separation in the commuting directions,
\eqn{lwedge}{
(x-y)^2_c = -(x^0 - y^0)^2 + \sum_{i \,\in C}  \, (x^i - y^i)^2  \ , 
\qquad {\rm where} \;\; C = \{ i  : \th^{ik} =0 \; \forall \; k \} \ . }
For example with $\th^{23} \neq 0$ we have $(x-y)_c^2 =  -(x^0 - y^0)^2 +  
(x^1 - y^1)^2$.  

This argument has two weaknesses: since it appeals to the violation of
Lorentz invariance, it would appear to apply whenever we have Lorentz
violating interactions, and not only in non-commutative field
theories. Also, the $SO(1,1)$ symmetry does not really determine a
light cone; while the light wedge used in \spmicro\ respects
$SO(1,1)$, so does the original light cone of \umicro. In
\cite{Chu:2005nb}, these weaknesses were addressed by performing an
explicit perturbative calculation of the commutator for \nc\ field
theory, showing that the micro-causality condition \spmicro\ is
correct for \snc\ field theories, whereas \umicro\ remains correct for
theories with local Lorentz violating interactions. This shows that
the non-local character of the \nc\ interactions is important.

We will give a brief derivation of this result
\cite{Chu:2005nb}. The authors consider \snc\ $\vp^3$ theory in
$\R^{5,1}$ for simplicity,
\eqn{ncphic}{
S = \int \, d^6 x \, \( - {1\over 2 } \, \p_\mu  \vp \star \p^\mu  \vp - {1\over 2} \,
m^2 \, \vp \star \vp  
+ {1\over 3!} \, g \, \vp \star \vp \star \vp  \) \ .}
The strategy is to calculate the expectation value of the Heisenberg 
picture field operator between two states $\ket{\a}$ and $\ket{\b}$, 
obtained by time-evolution of the perturbative vacuum $\ket0$, 
\eqn{fieldcom}{
\CM = \bra{\a}  \[ \vp_H(x_1) , \vp_H(x_2) \] \, \ket{\b} \ . }
Given a perturbation expansion of 
\eqn{compert}{
\CM = \sum_n \, g^n \, \CM^{(n)} \ ,}
one can study  $\CM^{(n)}(x)$ to determine the domains where 
it is guaranteed to vanish.  We have used translational invariance to write 
$x = x_1 -x_2$. This will provide us with the definition of the 
micro-causal light cone in field theory.

The calculation of $\CM^{(n)}(x)$ proceeds in the standard fashion
and some of the details can be found in \App{appA}. The main point 
is that we can write $\CM^{(n)}(x)$ in an integral representation as
\eqn{intrep}{
\CM^{(n)}(x) = N_n \, \[  \int  \, \prod_i \, d \l_i \, G_n(\l_i ,x) 
- \int  \, \prod_i \, d \l_i \, G_n(\l_i ,- x) \] \ , }
where the integrand $G_n(\l_i,x)$ generically takes the form
\eqn{intgnd}{
G_n(\l_i,x) = \exp \(
i \, { x^2 \over 2u_x} \, \sum_{i} \, \l_i  -i\, {m^2 u_x \over 2}\, 
 \sum_i \,  {1\over \l_i}   \) \, H_n(\l_i,x) \ . }
The integration variables $\l_i$ are the independent light cone momenta 
after imposition of momentum conservation and $N_n$
are constants. The details of the interaction are contained in the kernel 
$H_n(\l_i,x)$.  In writing \intrep, we have rewritten various amplitudes 
in terms of light cone variables for ease of computation:
$ x^2 = - 2 \, u_x \, v_x + {\vec x}^2 $.

The behaviour of the integrand $G_n(\l_i,x)$ in the complex space 
parameterised by $\l_i$ is crucial in determining the convergence properties 
of the integral. For instance, at tree-level in perturbation theory we have
\eqn{gzero}{
G_0(\l ,x) = {1 \over 4 \, \pi \, \l} \, \,\({i \, \l \over 2 \, \pi \,x } \)^2 \, 
\exp \( i \, \l {x^2 \over \, 2 \,u_x} - i \, {m^2 \, u_x \over 2 \, \l} \) \,
}
Given this, the usual contour deformation arguments can be used to show that 
\eqn{mzero}{
\CM^{(0)}(x) = 0 \;\;\;  \iff \;\;\;  x^2 = (x_1 - x_2)^2 > 0 \ ,} 
which is the expected result for the free field theory (recall that 
the \nc\ deformation affects quadratic terms only by the presence 
of irrelevant phase factors). To be precise, assuming $u_x >0$ without 
loss of generality, for $x^2 >0$ we can rotate the contour of integration 
for the $\l$ integral as $\l \to i \, \l$ in integrating $G_0(\l,x)$ 
(and $\l \to -i \, \l$ for $G_0(\l,-x)$).  The two integrals then are 
convergent and moreover the finite answers cancel in the difference; 
hence the commutator vanishes. For timelike separation $x^2 <0$,
there is no deformation of the contour that gives a convergent 
integral; hence generically the difference of the two integrals is 
non-vanishing and we are led to \mzero.

 For \nc\ field theories, we intuitively expect to see the first sign of 
 interesting effects at one-loop,  \ie, in $\CM^{(2)}$, for we 
 have qualitatively new feature in perturbation theory in the 
 form of non-planar Feynman graphs. We should expect that the 
 phase factors involved in the \nc\ $\vp^3$ interaction modify 
 the properties of $G_2(\l_i,x)$ and thereby deform the nature 
 of the light cone. Detailed calculations \cite{Chu:2005nb}
 show that this expectation is indeed
borne out. We in fact find that (see \App{appA} 
 for details)
\eqn{gtwo}{
G_2(\l_1,\l_2,x) \sim \exp \( i \, (\l_1 + \l_2 ) \, {  x_c^2 \over 2 \, u_x} \) \ , }
implying that 
\eqn{mtwo}{
\CM^{(2)}(x) = 0 \;\;\;  \iff \;\;\;  x_c^2 = (x_1 - x_2)_c^2 > 0 \ ,} 
with $x_c^2$ is as defined in \lwedge. This follows by repeating the 
argument for the convergence of the integrals with the modified 
integrand \gtwo.  So we see that the micro-causality 
condition in \snc\ field theories is modified at one-loop.  
In particular, note that the presence of the \nc\  interactions 
demand that the expectation value of the field commutator vanish 
only outside a light-wedge as surmised earlier\footnote{As an interesting 
aside, it should be possible to show the modification of the micro-causality
condition using the 1PI effective action derived for \nc\ $\phi^3$
theory in \cite{Minwalla:1999px}.  The inverse propagator (in momentum
space) for this theory is $p^2 + m^2 + g^2 \, h /p\circ p$, with
$p\circ q = p_\mu \, (\th^2)^{\mu \nu} \, q_\nu$. This propagator has
extra poles in the complex momentum plane at $p\circ p = -
h\,g^2/(p_c^2 +m^2)$, where $p_c$ is the restriction of the momentum
to the commutative sub-space. The presence of these poles induces a
delta function for the momentum along the \nc\ directions. This ultra
localization in momentum space is the origin of the light-wedge
\lwedge.}.

This micro-causal light wedge agrees with the prediction of the
spacetime in the previous section. This agreement should be considered
as qualitative evidence that the micro-causal structure in the field
theory is indeed related to the light cone determined by the bulk
spacetime.

\newsubsection{Micro-causality in \nc\ gauge theories}

We have shown that the micro-causality in a scalar \nc\ field theory
agrees with the bulk light cone. However, this scalar field theory is
only a toy model, and there are important differences which make it
unclear if this behaviour will generalize to the non-commutative gauge
theory which is actually dual to the bulk spacetime \spncmet. We will
now review the salient differences, and argue that we nevertheless
still expect to see a reflection of the bulk light cone in the
micro-causality of the \nc\ gauge theory.

The first difference has to do with the structure of UV divergences in
gauge theories. As remarked in the footnote above, the modification of
the micro-causality condition is related to the IR divergences seen in
perturbation theory for \nc\ dynamics. Although explicit calculation
confirms the presence of such IR poles in a general \nc\ gauge
theory~\cite{Matusis:2000jf}, they will be absent for the $\CN = 4$
\nc\ SYM theory we're interested in because of the
supersymmetry. Thus, the modification of the micro-causality condition
for gauge theories can't arise in the same way as it did in the
scalar theory studied above. 

There is however another important difference, which we will argue can
lead to a similar modification of the micro-causality condition by a
more subtle route. In \nc\ gauge theories there are no gauge invariant
local operators (essentially because translation in the \nc\ directions 
is equivalent to a gauge transformation), so it is unclear why we should even 
consider the micro-causal structure based on local fields. An over-complete set of
non-local gauge invariant observables was constructed
in~\cite{Ishibashi:1999hs, Gross:2000ba, Das:2000md}. 
The idea was to string a local operator such
as ${\rm Tr} (F_{\mu \nu} \, F^{\mu \nu})$ (which is of course gauge
invariant in the commutative limit) with an open Wilson line and take
its Fourier transform.  This defines a local gauge invariant operator
in momentum space $\CO(k)$.  They calculated correlation functions of
the operators $\CO(k)$ and showed that the correlation functions grow
exponentially in momenta
\eqn{ncymow}{ \langle \CO(k) \, \CO(-k) \rangle \sim \exp\(
\sqrt{{g_{YM}^2 \, N \over 4 \, \pi} \, |k \,\theta| \, |k| } \) \ . }
This exponential growth of correlation functions may also 
be seen from the supergravity dual~\cite{Gross:2000ba}. 
In writing the above we have dropped some regularization 
dependent terms which will be unimportant for properly 
renormalised correlators. We want to suggest
that the exponential growth of the correlation functions
signals a modification of the micro-causality
conditions for these operators. 

The clue comes from studies of a similar behaviour in another
non-local quantum field theory discovered in the past decade, little
string theory, which is a Poincar\'e invariant theory in six
dimensions with a mass scale $l_s^{-1}$ , living on the world-volume
of NS5-branes. These theories admit gauge invariant local operators in
momentum space and their correlation functions also grow exponentially
in momenta~\cite{Minwalla:1999xi}. This fact was used
in~\cite{Kapustin:1999ci} to argue that little string theories are
quasi-local theories, and that a modified notion of micro-causality
could be defined for such theories.

In local quantum field theories, correlation functions in position
space, the Wightman functions, are to be smeared with suitable test
functions to generate observables. We would therefore normally define
a micro-causality condition by smearing the Wightman function with
test functions of spacelike separated support. However, to do so, the
Wightman functions need to be tempered distributions, so that we can
use the usual Schwartz space of test functions. Physically this
amounts to the correlation functions in momentum space growing at most
polynomially.

In the non-local theory, where momentum space correlators grow
exponentially, the Wightman functions are rather singular
distributions and the allowed space of test functions is
restricted. In this case test functions are required to be real
analytic in position space, which precludes local observables (\cf
\cite{Kapustin:1999ci} for references and rigorous arguments). This
prevents us from defining a micro-causality condition in the usual
way, as we cannot smear with test functions of spacelike separated
support.

In~\cite{Kapustin:1999ci}, it was proposed that we can still study
micro-causality in these cases, by considering the analytic structure
of the Wightman functions at points where they make sense as
functions. Usually micro-causality implies permutation symmetry of
Wightman functions: for example, $W(x_1,x_2) = W(x_2,x_1)$ for
spacelike separated points $x_1$ and $x_2$. However, for the more
singular Wightman functions in a non-local theory, the region of
analytic behaviour is restricted. In the case of little string
theories,~\cite{Kapustin:1999ci} observed that the Wightman function
is analytic only for $\(x_1 -x_2\)^2 > l_s^2$, and proposed that
the micro-causality condition be replaced by requiring permutation
symmetry of Wightman functions in this restricted region.  The usual
commutativity condition is modified by the presence of poles in the
region $0< \(x_1 - x_2 \)^2 < l_s^2$.

While a detailed analysis of the analyticity properties of the \nc\
gauge theory Wightman functions is beyond the scope of the current
paper, we believe that it should be possible to construct an analogous
argument to determine the micro-causality properties associated with
the gauge-invariant operators $\CO(k)$, relating the restriction to
the light-wedge \mtwo\ seen from the spacetime point of view to the
fact that the momentum space correlator \ncymow\ is growing only along
the non-commutative directions. Recent discussions of microcausality 
conditions in non-commutative theories can be found in ~\cite{  Chaichian:2004qk, Franco:2004gx, Franco:2004fp}. 

It is also useful to note that agreement is expected: the
micro-causality conditions for an interacting theory on the boundary
ought to be captured by a free-field theory in the bulk in the large
$N$ limit (where the supergravity approximation is expected to valid),
because the correct micro-causality conditions in the field theory
stem from the quantum 1PI propagator. This might seem a bit puzzling
in the context of the previous discussion of \nc\ scalar field theories,
as we see (in appendix A) that the effect of non-commutativity arose
from non-planar diagrams in the field theory.  However, these
non-planar contributions are associated with the \nc\ interactions,
and are not suppressed by large $N$ power counting; in their absence,
the spacetime dual for \nc\ $\CN=4$ SYM would have been just $\A5S5$,
rather than the spacetime \spncmet\ with its complicated asymptotics.

\newsection{Generating spacetimes by null Melvin twist}
\label{nmt}

We next want to extend the discussion to the spacetimes dual to
lightlike \nc\ field theories. We will begin by presenting the
derivation of these solutions, using a particular solution generating
technique in supergravity, called the null Melvin twist
\cite{Alishahiha:2003ru, Gimon:2003xk}. This is slightly more
complicated than the twist used to obtain the geometry in the
spacelike case, and will add a B-field oriented along a lightlike
direction. While some of the solutions we are interested in have been
obtained previously in the literature \cite{Alishahiha:2000pu},
re-deriving them in this way will simplify the discussion of
generalisations. A nice summary of the solution generating scheme 
we use and some classifications of solutions can be found in 
~\cite{Hashimoto:2004pb}.
\newsubsection{The null Melvin twist}

The solution generating scheme is:

\noindent
{\bf 1.} Start with a solution of IIA/IIB supergravity,
with a non-compact translationally invariant direction; the corresponding 
Killing vector will be taken to be $\dya$.

\noindent
{\bf 2.} Boost the
geometry  in the $y$ direction by an amount $\gamma$. Note that this is 
just a coordinate change, which effectively adds momentum charge to the 
solution we start with.

\noindent
{\bf 3.} Now perform a T-duality along $y$, to get to a solution of IIB/IIA
supergravity. 

The above steps will typically add a fundamental string charge to the
solution we start with. However, if the Killing field $\dya$ can be
paired with a timelike Killing field $\dta$ (so that our starting
solution has $SO(1,1)$ isometry), then no charge is added. Instead
Step {\bf 2} is trivial as the geometry is boost invariant. Step {\bf
3} is then just a diagonal T-duality.

\noindent
{\bf 4.} {\it The Twist:} Assume that we have in addition to the 
Killing field $\dya$, some other rotational or translational isometries. 
We would like to perform a non-diagonal T-duality by combining these 
isometries with that generated by $\dya$.
Schematically, denoting the one-forms dual to the additional isometries 
by $\sigma$, we perform a twist (a coordinate transformation) by replacing
\eqn{twist}{
\sigma \rightarrow \sigma + 2 \, \a \, dy \ .
}
Here $\a$ parameterises the amount of twisting. 

\noindent
{\bf 5. }We now T-dualize the geometry back to IIA/IIB along $y$. 
The twist followed by the T-duality is effectively a non-diagonal
T-duality.

\noindent 
{\bf 6. } Boost the solution by $-\gamma$  along $y$. The purpose of this
boost is in  part to undo the original boost performed.

\noindent
{\bf 7.} Now, we perform a double scaling limit, wherein the boost
$\gamma$ is scaled to infinity and the twist $\a$ to zero keeping
\eqn{doublescaling}{
\b  = {1 \over 2} \, \a \, e^\gamma = {\rm fixed} \ .}

The null Melvin twist transformations in steps 4 through 7 can be
thought of as converting the string solution into a fluxbrane,
followed by a boost and scaling to end up with a null
isometry.

\newsubsection{Null Melvin twist of the D3-brane solution}

Let us apply this transformation to the D3-brane geometry.  The
extremal D3-brane geometry is a solution of Type IIB supergravity with
metric:
\eqn{dtbrane}{
ds^2 = {1\over \sqrt{H(r)}} \, (-dt^2 + dy^2 + dx_1^2 + dx_2^2 ) + \sqrt{H(r)}
\, \( dr^2 + r^2 \, d\Omega_5^2 \) \ , }
with 
\eqn{Hdef}{
H(r) = 1 + {R^4 \over r^4} \ . }
The metric \dtbrane\ is supported by a five-form flux. Since the flux
turns out to be insensitive to the null Melvin twist we will not write
it explicitly.

The geometry \dtbrane\ has a $SO(1,1)$ Lorentz symmetry along the
world-volume directions $(t,y)$, and so we will not have any charge
added during the first two steps. The twisted T-duality implemented in
steps 4 and 5 requires us to pick a direction to do the twisting. The
simplest choice turns out to be the translationally invariant
directions $\dx1$ and $\dx2$.  Performing the twist as
\eqn{dttwista}{
dx_1 \to \; \; dx_1 + \a \; dy \ , \qquad dx_2 \to 
\; \; dx_2 + \a \; dy  \ , }
we obtain the following geometry: 
\eqn{dtnmtw}{\eqalign{
ds^2 &= - \, \b^2 \, H^{- {3 \over 2}}\,
 \(dt + dy\)^2 + {1 \over \sqrt{H}} \, \( 
-dt^2 + dy^2 + dx_1^2 + dx_2^2 \) +  \sqrt{H}
\, \( dr^2 + r^2 \, d\Omega_5^2 \) \ , \cr
B &= \b \,{1 \over H} \, (dt + dy) \,\wedge \, \( dx_1 + dx_2 \) \ . 
}}
Apart from the NS-NS 3-form flux written above the metric is in
addition supported by a five-form flux (which is the same as for the
original D3-brane solution \dtbrane). The sequence of operations
{\bf 1} to {\bf 7} maps a constant dilaton solution back to a
constant dilaton solution for this particular case.  Note that the
solution \dtnmtw\ is asymptotically flat. 

In deriving \dtnmtw\ we have used the translational isometries
$\dx{i}$ to perform the twisted T-duality. We could just as well have
used some angular isometries, such as the rotation isometry in the
$(x^1, x^2)$ plane, or isometries of the $\S^5$.  In such cases, we would
generate asymptotically plane wave, rather than asymptotically 
flat, spacetime; however, since this asymptotic region will be lost once 
we take the decoupling limit, the interesting causal features of the 
plane wave (such as 1-dimensional boundary or non-global hyperbolicity)
 would not enter our story.
These cases are
briefly discussed in \App{appB}, although the essential physical point
of interest is exemplified adequately by \dtnmtw.

\newsubsection{Near-horizon geometry of null Melvin twisted D3-brane}

Our main interest is to identify interesting generalisations of the
AdS/CFT correspondence, so we now consider an appropriate decoupling
limit of the geometry \dtnmtw, to obtain a new duality relating the
low energy degrees of freedom on the twisted D3-brane to a
supergravity geometry. For the D3-brane geometry \dtbrane, we know
that the near horizon limit corresponding to decoupling the closed
string modes from the open string modes on the D3-brane is obtained by
dropping the $1$ in the harmonic function $H(r)$ given in \Hdef
\cite{Maldacena:1997re}.  For the solution \dtnmtw, we want to ensure
that the effect of the twist survives in the decoupling limit.  Thus
the appropriate limit is analogous to the Seiberg-Witten scaling for
\nc\ field theories \cite{Seiberg:1999vs}. This again amounts to
dropping the $1$ from the harmonic function $H(r)$ \Hdef, now in the
new metric \dtnmtw. The resulting geometry is given by
\eqn{dtnmnhor}{\eqalign{
ds^2 &= - \, \b^2 \, {r^6 \over R^6} \,du^2 + {r^2 \over R^2}\, \( 
-2 \, du \, dv  + dx_1^2 + dx_2^2 \) +  {R^2 \over r^2} 
\, \( dr^2 + r^2 \, d\Omega_5^2 \) \ , \cr
B &= \beta \,{r^4 \over R^4} \, du \,\wedge \, \( dx_1 + dx_2 \) \ ,  
}}
where we have introduced light cone coordinates $u = t +y$ and $ v =
t-y$. In addition to the NS-NS B-flux there is also a five-form flux,
which is identical to that supporting the $\A5S5$
solution. The metric only differs from the $\A5S5$ metric written in
Poincar\'e coordinates by the term proportional to $\b^2$.  
This solution was previously derived in \cite{Alishahiha:2000pu}.
They considered the solution to Type IIB supergravity that was known
to be dual to $\CN= 4$ SYM deformed by spacelike non-commutativity
\cite{Hashimoto:1999ut, Maldacena:1999mh}, and boosted it to derive
\dtnmnhor. As in the case of spacelike non-commutativity, it is difficult to
define a conformal boundary for this spacetime. The correspondence to
the field theory is derived by thinking of the field theory as living
on the D3-branes that source the full geometry \dtnmtw, and taking the
low-energy limit.

We now show that this solution is non-singular, and
non-distinguishing. We show that it is non-singular by observing that
all the curvature invariants for \dtnmnhor\ are the same as for $\A5S5$, 
and that the spacetime is geodesically complete. The
additional contributions to the curvature from the term proportional
to $\b^2$ will involve $\dua$, so they will not change the curvature
invariants. Geodesic completeness can be shown either by observing
that the metric is conformal to a pp-wave,
\eqn{ndgeo}{\eqalign{
ds^2 &=  {r^2 } \, \( -2 \, du \, dv  - \b^2 \, r^4  \, du^2+ 
dx_1^2 + dx_2^2 +  {dr^2 \over r^4} + {1\over r^2} \, d\Omega_5^2 \) \ , \cr
& = {1\over z^2} \, \(-2 \, du \, dv  - {\b^2 \over z^4}  \, du^2
+ dx_1^2 + dx_2^2 +  dz^2 +  z^2 \, d\Omega_5^2 \) \ 
}}
(where we have set $R=1$ and in the second line used $ r = 1/z$), 
and appealing to the results of~\cite{Hubeny:2002zr}, or by
explicitly studying the geodesics. We will discuss timelike geodesics 
in the geometry \ndgeo\ in \sec{cll}. The essential point 
is that the motion in the radial direction is confined to 
a finite range and inertial observers are unable to escape 
towards the asymptotic region  (\cf, \req{lgeodtll}, where it is 
clear that geodesics cannot access the region $z =0$, 
or equivalently $ r = \infty$, due 
to a potential barrier). Thus, though we have diverging curvatures in 
the large $r$ region, the spacetime is non-singular.

\begin{figure}[htbp]
\begin{center}
\includegraphics[width=6in]{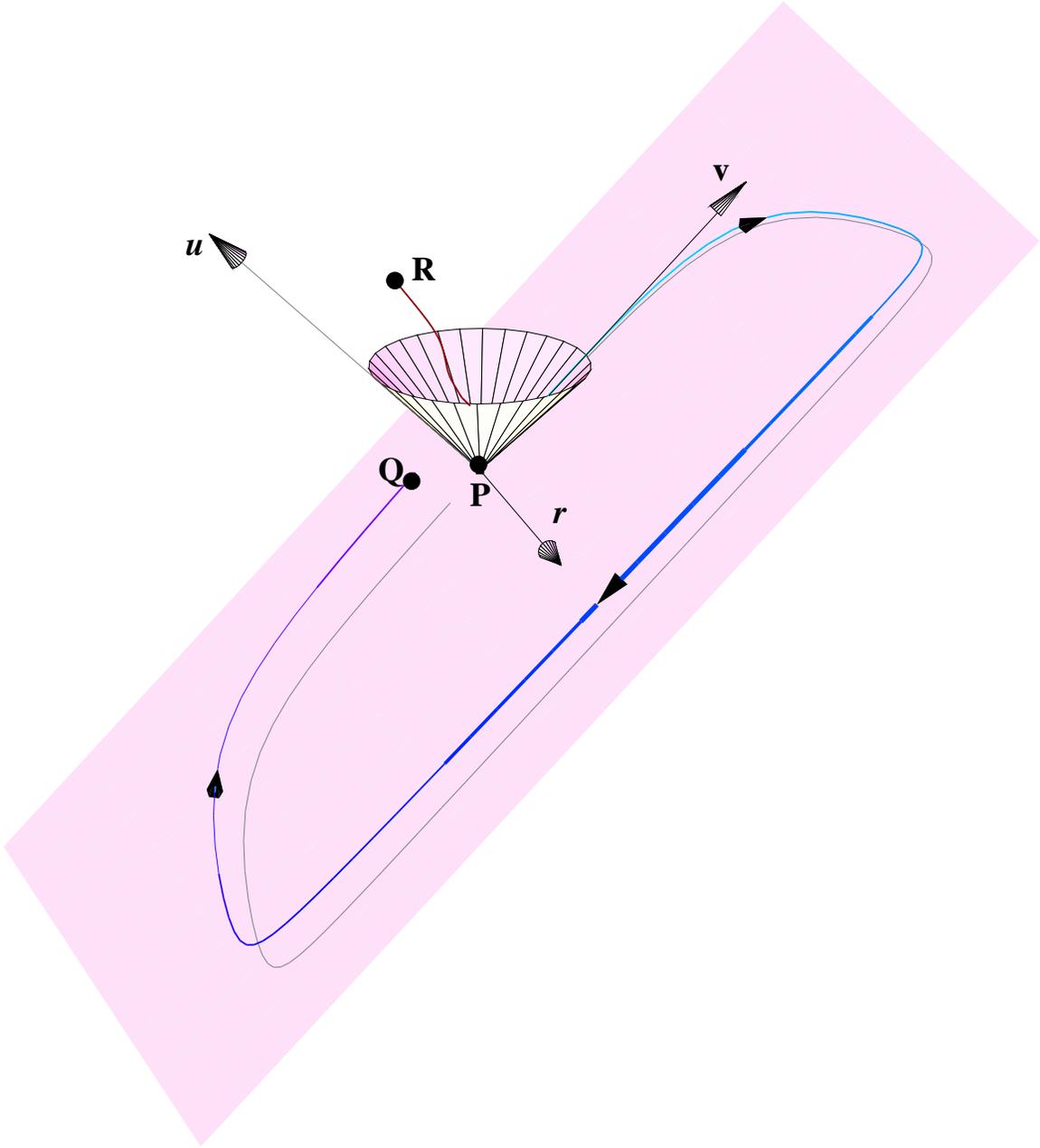}
\caption{{\small Light cone structure in the non-distinguishing spacetime
\ndgeo. From 
point ${\bf P}$ we can reach point ${\bf R}$ by a usual causal curve. To reach
point ${\bf Q}$ which is spacelike separated in the Minkowski light cone, we 
need to use the asymptotic structure of the spacetime. By using curves 
that venture out further in $r$ we can arrange for the causal curve to return 
arbitrarily close to the starting point. As a result the future of the point ${\bf P}$ 
is the part of the spacetime above the constant $u$ plane shown above.  }}
\label{nondistlc}
\end{center}
\end{figure}

To show that the spacetime is non-distinguishing, it is easiest to use the
representation \ndgeo\ of the metric as conformal to a pp-wave, and
use the results of~\cite{Flores:2002fx, Hubeny:2003sj}. Since causal
properties are invariant under conformal transformations, the causal
character of our spacetime is determined by that of the pp-wave
spacetime within the parenthesis in \ndgeo.  In \cite{Flores:2002fx,
Hubeny:2003sj} it was shown that these pp-wave spacetimes are
non-distinguishing.  Causal curves that connect a point $P= (u_0, v_0,
x_0^i, z_0)$ in the spacetime to any point $Q = (u_0 +\e, v_1, x_1^i,
z_1)$, with $\e >0$ and arbitrary values of $(v_1,x^i_1, z_1)$ were
explicitly constructed. This shows that the causal future of $P$ is
the entire region $u \ge u_0$, as depicted in \fig{nondistlc}.  In the
terminology of \cite{Flores:2002fx}, the metric \ndgeo\ written in the
$r$ coordinate is conformal to a super-quadratic pp-wave, as $g_{uu}$
grows faster than a quadratic for large $r$.  We give an explicit 
construction of the causal curve from $P$ to $Q$ (in the $r$ coordinates) in \App{appC}.

 \newsection{The holographic dual field theory}
\label{ll} 

As can be seen either from the lightlike nature of the B-field in the
spacetime geometry \dtnmtw\ or from the alternative derivation\footnote{
In \cite{Alishahiha:2000pu}, the solution was obtained by
starting from the holographic dual to NCYM with spacelike
non-commutativity and then boosting.
} of this geometry 
in \cite{Alishahiha:2000pu}, the dual field theory description
of the geometry \ndgeo\ is a $d=4$, $\CN =4$ non-commutative Super
Yang-Mills (NCYM) on $\R^{3,1}$ with a constant lightlike
non-commutativity parameter $\theta^{v \, i} = \beta$ for $i = 1,2$.

\newsubsection{Well behavedness of observables in the field theory}

Such lightlike non-commutativity was first discussed in
\cite{Aharony:2000gz}, where it was shown that theories with
lightlike non-commutativity behave like field theories with non-local
interactions. This is in contradistinction with the case of timelike
non-commutativity, where we have a string theoretic behaviour due to
the non-decoupling of excited open string oscillators. The crucial
difference between the two cases is the following~\cite{Aharony:2000gz}: 
in the case of timelike non-commutativity, we
start from a D-brane background and turn on electric fields on the
brane world-volume~\cite{Seiberg:2000ms}. There is however a critical
limit to the electric field (T-dual to there being a maximum velocity
-- speed of light), and at the critical point the open strings become
effectively tensionless.  The timelike theory is defined by zooming in
on this critical region, and as a result one ends up with a finite
effective string tension. In the case of lightlike non-commutativity,
we turn on both electric and magnetic fields, and it is easy to see
that this combination doesn't lead to a critical electric
field. Alternatively, one can think of the lightlike theory as a
boosted version of the spacelike theory, where there is clearly no
critical behaviour.  Thus, this dual is a well behaved field theory
with non-local interactions.  Hence it is quite surprising that the
dual geometry \ndgeo\ exhibits some causal pathology.

To understand better the physics of \lnc\ field theories we can
consider observables in these theories, such as the correlation
function of local operators $\CO(p)$. These can be calculated at
strong 't Hooft coupling using the using the usual bulk-boundary
correspondence. It transpires that massless minimally coupled scalar
fields in the bulk geometry \ndgeo\ satisfy a Mathieu equation which
is very similar to that obtained in~\cite{Maldacena:1999mh} for the
geometry \spncmet.  Of course the parameters appearing in the equation
have a different dependence on momenta, as should be expected given
the different symmetries of the geometries. This implies that the
correlation functions are schematically similar to the spacelike
case. Likewise, one can consider gauge invariant observables built
with open Wilson lines as discussed in~\cite{Gross:2000ba} and check
that these are also well behaved and in fact will be exponential 
in the momenta. In summary, there is no reason from
a field theory perspective that \lnc\ field theories should be
pathological, which makes the bizarre causal structure of the dual
spacetime geometry all the more interesting.

\newsubsection{Genericity of non-distinguishingness for \lnc\ field theories}

We have shown above that the supergravity background dual to 
\lnc\ $\CN =4 $ SYM is a non-distinguishing
spacetime. We will now show that any well defined field 
theory will produce a non-distinguishing holographic dual when we 
deform it by adding some lightlike non-commutativity.

Consider any field theory defined in the usual Wilsonian sense, by its
ultraviolet modes. From the UV/IR correspondence for gauge/gravity
duality~\cite{Susskind:1998dq}, we know that the field theory
ultra-violet corresponds to the infra-red region of the supergravity
dual. For field theories on $\R^{3,1}$ one can without loss of
generality assume that the holographic dual geometry is of the
following warped product form:
\eqn{dualgeo}{
ds^2 = A(r)^2 \, \(-dt^2 + dy^2 +dx_1^2 + dx_2^2 \) + dy_6^2 \ ,
}
where $r$ is an effective radial coordinate on the transverse space
with metric $dy_6^2$.  If the field theory approaches a conformal
fixed point in the UV, the large $r$ behaviour will be asymptotically
${\rm AdS}_5 \times X^5$ for some compact Einstein manifold $X^5$.
Our argument will however be general enough to include more exotic
field theories which do not arise from nice conformal fixed points in
the UV, such as the Klebanov-Strassler cascade~\cite{Klebanov:2000hb}. 

We would like to turn on lightlike non-commutativity in the field
theory. We have seen above that this is achieved by a Null Melvin
Twist on the dual geometry \dualgeo.  The Poincar\'e invariance of the
field theories on $\R^{3,1}$ implies that there are no fluxes that
prevent us from carrying through the steps involved in the duality
chain. A null Melvin twist of \dualgeo\ will lead to the following
geometry:
\eqn{nmdualgeo}{
 ds^2 = A(r)^2 \, \(-2 \, du \, dv - \b^2 \, A(r)^4 \, du^2 
 +dx_1^2 + dx_2^2 \) + dy_6^2 \ .
}
The question is then what the effect of the additional $du^2$ term is
on the causal structure; this depends on the particular form of
$A(r)$. For field theories with a conformally invariant fixed point in
the UV, the large $r$ behaviour of $A(r)$ is $A(r) \sim e^{a r}$,
giving the same kind of asymptotics as in the particular case we
studied, and hence implying that for such cases, the metric \nmdualgeo\ is
non-distinguishing. The same is true for the $A(r)$ appropriate to the
Klebanov-Strassler geometry~\cite{Klebanov:2000hb}, so \nc\
deformations of this theory also have non-distinguishing
duals, see \App{appB} for the explicit metric.

Thus, we see that the UV behaviour of the field theory is responsible
for the dual supergravity background being causally ill-behaved. It
follows that infrared modifications of the field theory will not
remove the causal pathologies, as they do not change the asymptotics
of the dual spacetimes. For example, the thermal version of the \nc\
theory will be dual to a spacetime with a black hole in it, but this
will still be non-distinguishing.  For completeness, we present the
metric for the thermal version of the \lnc\ $\CN=4$ SYM in 
\App{appB}.

\newsection{Causality in \lnc\  theories}
\label{cll}

We now attempt to understand the origin of the non-distinguishing
character of the spacetime \ndgeo\ from the field theory perspective.
As with the case of spacelike non-commutativity, it will be
instructive to first understand the behaviour of the bulk light cone
in the asymptotic region.  We will then proceed to look at the
micro-causality condition in field theory and show that the
perturbative micro-causal light cone is modified so as to be
consistent with the characteristics of the bulk spacetime. As in
section 3, we consider a simple scalar field theory as a model, but it
seems reasonable to expect that this micro-causal structure will be
independent of the details of the particular field theory we consider.

\newsubsection{Lightlike non-commutativity: bulk light cone}

Our derivation of the bulk light cone for the \lnc\ field theories will 
proceed in a fashion analogous to the spacelike case discussed in 
\sec{splw}. We will focus once again on the properties of the bulk 
light cone in the asymptotic region of the spacetime, which for 
\ndgeo\ will be the region $r \to \infty$ or $z \to 0$. 

Let us consider the induced metric on a surface of fixed $r = r_0$, 
so as to discern the causal properties in the cut-off \lnc\ field theory. 
This induced metric is (in what follows we will ignore the $\S^5$ directions)
\eqn{indmetll}{
ds^2 \approx r_0^2 \, \( - 2 \, du \, dv  - \b^2 \, r_0^4 \, du^2 +
 dx_1^2 + dx_2^2 \) \ , }
which of course is the metric on flat $\R^{3,1}$ once we rescale the
coordinates appropriately. However, if we restrict to homogeneous
rescaling of the coordinates, the metric \indmetll\ degenerates to a
one dimensional metric due to the dominance of $g_{uu}$ for large
$r_0$. We then expect the bulk light cone to degenerate to a Galilean
causal structure, in which any two points are causally related unless
$\Delta u = 0$.

The above argument for the asymptotic bulk light cone can be made
precise by considering the two-point function. For the metric \ndgeo,
the free scalar wave equation is still the Mathieu equation, making
explicit determination of the propagator tricky. We will therefore
concentrate again on a geodesic approximation to the propagator. Let
us begin by considering timelike geodesics in the geometry \ndgeo,
\eqn{tgeoll}{
-1 = {1\over z^2} \, \( - 2 \, \ud \, \vd - {\b^2 \over z^4} \, \ud^2 + \xd1^2 
+ \xd2^2  +\zd^2 \) \ , }
where $ ^{.} \equiv {d \over d \t}$, with $\t$ the affine parameter
along the geodesics.  The Killing symmetries $\dua$, $\dva$ and $\dx{i}$
determine conserved energies and momenta $(p_u, p_v, p_i)$,
respectively. The geodesic motion reduces to a classical problem for
the zero energy trajectories of a particle in an effective potential%
\footnote{From the form of $\Veff(z)$ it is clear that geodesics never 
reach $z =0$ for $p_v \neq 0$, thereby preventing inertial observers 
from being subject to large tidal forces resulting from the diverging 
curvatures in that region. This is the hitherto alluded to characteristic 
that demonstrates geodesic completeness of \ndgeo.}
$\Veff(z)$:
\eqn{lgeodtll}{\eqalign{
 \zd^2 + \Veff (z) &= 0, \cr
\ud  = -p_v \, z^2 \ , \qquad & \vd = - p_u \, z^2 + {\b^2 \, p_v \over z^2} \ , 
\qquad \xd{i} =  p_i \, z^2 \ , \cr
\Veff(z)   = z^2 +&\(p_1^2  + p_2^2 -  2 \, p_u \, p_v \) \, z^4 + p_v^2 \, \b^2  
.}}
Let us consider geodesics that travel an infinitesimal distance in the
radial ($z$) direction.  Choose $ z = z_0 \, \( 1 +\e \)$ with $\e \ll 1$
such that $\Veff(z_0) =0$. Then we have
\eqn{infgdsll}{
z_0^2 \, \ed^2  \approx 2\,  z_0^2 \, \e \( 1 + 2 \, {\b^2 \,p_v^2  \over z_0^2 } \) 
\equiv 2 \, \e \, \a .} 
This allows us to approximately integrate the geodesic equations to
obtain, 
\eqn{intdtll}{
\D u=p_v \, z_0^3 \, {\sqrt{2\, \e}\over \sqrt{\a}} \ , \qquad 
\D v=\(p_u - { \b^2 \, p_v \over z_0^4} \) \, z_0^3 \, {\sqrt{2\, \e}\over \sqrt{\a}} \ ,
\qquad 
\D x_i=p_i \, z_0^3 \, {\sqrt{2\, \e}\over \sqrt{\a}} \ ,
 }
It is obvious from \intdtll\ that one can make the distance travelled by the 
geodesics in individual directions large by choosing appropriate momenta. 
However, the motion is confined to the region in the immediate vicinity 
of the classical bulk light cone, as
\eqn{lindis}{
-2 \, \D u\, \D v -{\b^2 \over z_0^4} \, \D u^2+ \D x_1^2 + \D x_2^2  
\approx -{2\, \e \over \a} z_0^4 \ll 1  \ .
}
Thus, bulk geodesics can relate points inside the effective light cone given by
\eqn{lblwedge}{
- 2 \, \D u\, \D v -{\b^2 \over z_0^4} \, \D u^2+ \D x_1^2 + \D x_2^2 = 0,
}
which limits as $z_0 \to 0$ to the Galilean invariant light wedge. Causal properties
are then determined by just the value of $u$ coordinate. We see that 
the point $P = (u_0, v_0, x^i_0)$ is in the past of $Q = (u_1, v_1, x^i_1)$ 
if $u_0 < u_1$ irrespective of the other coordinates.

\newsubsection{Micro-causality in \lnc\ field theories}

Let us now study the micro-causal light cone for this case of
lightlike non-commutativity, following the discussion of the
spacelike case given earlier. First of all, we can use the simpler
argument of~\cite{Alvarez-Gaume:2003mb} to suggest the appropriate
micro-causality condition for this case.  Suppose we have a field
theory in $\R^{d,1}$ with $\th^{u i} \neq 0$, where we use lightcone
coordinates $(u,v,x^i)$ for $i = 1, 2, \cdots, d-1$. This \nc\
deformation will break the Lorentz group $SO(d,1)$ down to a Galilean
group. The natural Galilean invariant micro-causal relation to consider
is
\eqn{lmicro}{
\langle \, [\Phi(x), \Phi(y)]_{\pm} \, \rangle 
= 0 \, \qquad { \rm for } \;\; u_x =  u_y \ . }

We can use the  argument of~\cite{Chu:2005nb}  as briefly
reviewed in \sec{bdysp} to show that the \lnc\ field theory in 
fact has such a micro-causality condition, while a local quantum 
field theory with Galilean invariance 
will still have the usual commutation relations \umicro.

For sake of simplicity we  consider  the \nc\ $\vp^3$ theory introduced 
in \ncphic; we will calculate the perturbative 
behaviour of the field commutator expectation value to 
probe the micro-causality condition. In fact, the calculation 
proceeds in a manner similar to the \snc\ case, but for a few essential 
differences in the details of the kernels. As in that case there is no 
difference from commutative $\vp^3$ theory at the tree level \ie, for 
$\CM^{(0)}$. Once again interesting effects show up at the one-loop
contribution to the commutator, thanks to the contribution from 
the non-planar diagrams. The calculation is outlined in \App{appA}.
We find that the non-planar graphs give us one-loop contribution 
\eqn{gtwol}{
G_2^{(np)}(\l_1,\l_2,x) \sim \int_0^\infty \,  d\l_1 \, d\l_2 \, \exp \( i {{\vec \th}\cdot{\vec \th}
\over 2 \, u_x } \, \l_1 \, \l_2 \, \(\l_1 + \l_2\)  - i \, {m^2 \, u_x \over 2} \, 
\({1\over \l_1} + {1\over \l_2} \) \) \ . }
In writing the above we have retained only the dominant contribution at 
large momentum (for this is the part that determines the convergence properties)
and ${\vec \th}$ is a vector with components $\th^{ui}$.
We have furthermore
\eqn{mtwol}{
\CM^{(2)}_{np}(x) \sim \int_0^\infty \, d\l_1 \, \int_0^\infty \, d\l_2 \, 
G_2^{(np)}(\l_1,\l_2,x) - \int_0^\infty \, d\l_1 \, \int_0^\infty \, d\l_2 \, 
G_2^{(np)}(\l_1,\l_2,-x) \ , }
and as usual, $\CM^{(2)}_{planar}(x)$ is the same as in the commutative 
field theory. 

In order to ascertain the micro-causality condition, we need
to figure out the domain where $\CM^{(2)}_{np}$ is guaranteed to
vanish. This can be done by looking at the analytic properties of
$G_2^{(np)}(\l_1,\l_2,x)$. Assuming without loss of generality $u_x >
0$, we see that $G_2^{(np)}(\l_1,\l_2,x)$ will be convergent only for
$\l_{1,2} \to -i \, \infty$. However, for $\l_{1,2} \to -i \, 0$ we
encounter a divergence from the term proportional to $m^2$ in the
exponential. Hence we are forced to conclude that there is no domain
in the complex $\l_{1,2}$ plane where the integral of $
G_2^{(np)}(\l_1,\l_2,x)$ is absolutely convergent.  This in particular
implies that we are generically going to encounter a non-vanishing
value of $\CM^{(2)}$. The only special case is when $u_x =0$ when we
expect $\CM^{(2)}$ will indeed vanish. This is not apparent from
\gtwol\ and \mtwol, since these are written in a light cone
quantization scheme. Intuitively, one expects this to arise simply from
the fact that this is indeed the `equal time' commutation relation for
light cone quantization.  To summarise, the micro-causality condition
for this \lnc\ field theory is indeed as given by \lmicro.

As we have emphasized in the \snc\ case, it would be challenging to
extend this computation to the non-commutative Yang-Mills theory which
is actually dual to the bulk spacetime. Once again, the appropriate 
framework for exploring these ideas would be to look at the details 
of the analytic behaviour of the Wightman functions. However, as argued 
in the spacelike case, the micro-causal light cone for the field theory ought 
to be in agreement with the causal properties of the supergravity dual. So,
it is again encouraging to find that the results in the scalar field
theory exhibit qualitative agreement with the bulk
spacetime. Since the bulk behaviour was intimately associated with the
non-distinguishing character of this spacetime, we might even say that
this non-distinguishing character is not only consistent in string
theory, but in fact necessary to reproduce this behaviour in the dual field
theory.

\newsection{Discussion}
\label{discuss}

We have studied modifications of the causal structure of the bulk
spacetime in the AdS/CFT correspondence, by considering the changes in
the geometry dual to a non-commutative deformation of the field
theory. We pointed out that these deformations cause radical changes
in the bulk causal structure. For spacelike non-commutativity, the
light cone in the boundary directions asymptotically becomes
independent of the non-commutative directions. For lightlike
non-commutativity, the modification is even stronger, producing a
non-distinguishing spacetime, in which points are always connected by a causal
curve unless they have the same value of `light-cone time'. 

We compared the asymptotic bulk light cone to the micro-causal light
cone in a scalar non-commutative field theory, following
\cite{Alvarez-Gaume:2003mb,Chu:2005nb}.  In this toy example we were
able to show that the micro-causal light cone was in complete
agreement with the bulk predictions. As we have already remarked, we
think it reasonable to expect this agreement to extend to $\CN =4$
\nc\ gauge theory, and have suggested a possible way to infer this
rigorously.

The essential issue in considering \nc\ gauge dynamics is the absence
of local gauge invariant observables in position space. However, any
bulk calculation of correlation function in strongly coupled \nc\
gauge theory will exhibit the deformed light cone structure. From the
point of AdS/CFT correspondence it is imperative that this structure
be reproduced from the field theory side as well. In fact, as
discussed in~\cite{Gross:2000ba, Rozali:2000np} there is an excellent agreement
between the result for the correlation function of gauge invariant
open Wilson loop operators in the field theory (obtained by summing up
ladder diagrams) and the bulk prediction.  Furthermore, given the
nature of observables in the \nc\ gauge theory, this agreement is
expected to be universal. Since the bulk two point function is
sensitive to the deformation of the light cone, it seems quite natural
to expect the same in the field theory.

These results provide a new connection between the causal structure in
the bulk and in the boundary. However, the full encoding of the bulk
causal structure in the dual field theory description remains an
important open problem. In particular, we have not addressed the very
interesting question of how the bulk causal structure is encoded when
the Poincar\'e invariance in broken only in the interior (as in black
hole solutions).  

The discovery that a non-distinguishing spacetime is related via the
AdS/CFT duality to a well-defined quantum field theory is of interest
in its own right. It suggests that such geometries may be more
respectable as string theory backgrounds than one might have
expected. The usual objection to non-distinguishing backgrounds, that
a small deformation can convert them into a solution with closed
timelike curves, is here circumvented by the fact that the
non-distinguishing character is a consequence of the behaviour of the
light-cone in the asymptotic region, where corrections are
suppressed. That is, although there are almost closed timelike curves,
they do not remain within a compact region of the spacetime. These
solutions deserve further investigation.

\section*{Acknowledgements}

 We would like to thank Henriette Elvang, Sean Hartnoll, Gary Horowitz 
 and especially Chong-Sun Chu for illuminating discussions. 
 VH and MR are supported by the  funds from the Berkeley Center for
Theoretical Physics, DOE grant DE-AC03-76SF00098 and
the NSF grant PHY-0098840. SFR is supported by the EPSRC Advanced 
Fellowship. 

\startappendix
\Appendix{Derivation of light-wedge}
\label{appA}

We present in this appendix  a short derivation of the micro-causality 
condition in \nc\ field theories. Further details can be found in 
\cite{Chu:2005nb}.

As discussed in \sec{bdysp}, we are interested in calculating the matrix 
element of the field commutator in perturbation theory. The field theory we 
consider will be \nc\ $\vp^3$ theory in $\R^{5,1}$ with Lagrangian \ncphic.
In what follows bold face letters will denote spatial vectors $\{\bp, \bx\}$ \etc, 
in $\R^5 \subset \R^{5,1}$. We will also have use for light cone coordinates 
and the vectors transverse to the light cone directions will be denoted as 
$\{\vec{p},\vec{x}\}$.

Working within the framework of canonical quantization, one can 
write the interaction picture field $\vp(x)$ as 
\eqn{intpic}{
\vp(x) = \int \, {d^5 \bp \over (2 \, \pi )^5 } \, {1 \over \sqrt{2 \, \om_p} }\, 
\(a_p \, e^{- i \, p . x } + a_p^\dagger \, e^{i\, p.x } \) \ , }
where $\om_p = \sqrt{\bp^2 + m^2}$.

As usual it is possible to pass between the interaction and Heisenberg pictures:
\eqn{passage}{
\vp_H(x) = U^\dagger (t,t_0) \, \vp(x) \,U(t,t_0) \ , \qquad 
U(t_1,t_2) = {\rm T} \, \exp \( i \, {g \over 3!}  \, \int_{t_1}^{t_2} \, dt \, \int \, 
d^5 x \, \vp \star \vp \star \vp  \) \ . }
We will make use of this to evaluate the matrix element \fieldcom\ and consider 
the states $\ket{\a}$ and  $\ket{\b}$ obtained by evolving the perturbative 
vacuum $\ket0$ by the evolutions operator $U^\dagger(t_1,t_0)$.

For the $\vp_\star^3$ theory the quantities of interest are the tree-level term $\CM^{(0)}$ 
and the one-loop term $\CM^{(2)}$. The former is just the usual free field 
propagator; the presence of \nc\ interactions play no interesting role for 
the quadratic terms in the Lagrangian \ncphic. In fact, we have
\eqn{mzeroa}{
\CM^{(0)}(x = x_1 -x_2) = \int {d^5 \bp \over (2 \, \pi^5)} \, {1\over 2 \, \om_p} \,
 e^{-i \, p.x} 
- \int {d^5 \bp \over (2 \, \pi^5)} \, {1\over 2 \, \om_p} \, e^{+i \, p.x}\ ,} 
which may be written in the form \intrep, using the identity
\eqn{identy}{
\int\, {d^5 \bp \over \om_p } = \int_0^\infty \, {d p^+ \over p^+} \, d^4{\vec p} \, 
\mid_{p^- =  {{\vec p}^2 + m^2 \over 2 \, p^+ } \ . }}
In writing the above we have passed over to light cone representation of 
momenta for convenience. Hence we obtain 
\eqn{mzerob}{
 \int {d^5 \bp \over (2 \, \pi^5)} \, {1\over 2 \, \om_p} \, e^{-i \, p.x}  = 
  2 \, (2 \, \pi)^5 \, \int_0^\infty \, {d p^+ \over p^+} \, d^4{\vec p} \, 
  \exp \( i\,  p^+ \, v_x +{{\vec p}^2 + m^2 \over  2 \, p^+ } \, u_x - 
  i \, {\vec p} . {\vec x} \) \,  }
Completing the Gaussian integrals over ${\vec p}$ 
in \mzerob\ we find the integral representation of $\CM^{(0)}$ with the 
kernel $G_0( p^+=\l,x)$ as given in \gzero. 

From this representation of $\CM^{(0)}(x)$ one can ascertain the domains 
where the commutator is required to vanish. As discussed in \sec{bdysp} 
the strategy is to ascertain the possible contour rotations for the $p^+ = \l$ 
integral. Depending on the convergence of the integrals of $G_0(\l ,\pm x)$,  
the two terms contributing to $\CM^{(0)}(x)$ mutually cancel. This is where 
we see the origin of the micro-causality condition in the field theory. The 
contour rotations that lead to cancellation of the two integrals work only for 
spacelike separated points and not for points $x_{1,2}^\mu$ which are 
in causal contact.

\begin{figure}[htbp]
\begin{center}
\includegraphics[height=1.8in,width=6in]{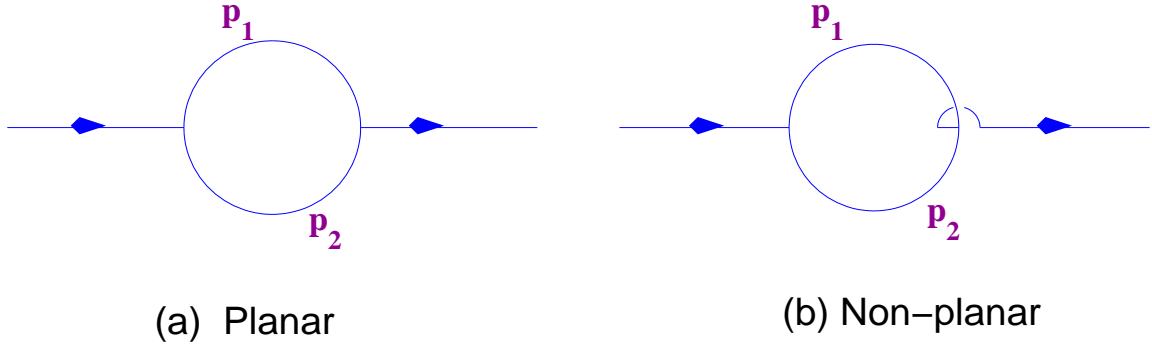}
\caption{{\small Feynman diagrams contributing to $\CM^{(2)}$. Due to the 
non-commutative interaction we have both planar and non-planar graphs. }}
\label{feyndiags}
\end{center}
\end{figure}

The calculation of $\CM^{(2)}$ proceeds in a similar fashion. We get contributions 
for the \nc\ $\vp^3$ theory from two kinds of one-loop diagrams: planar 
and non-planar. The planar diagrams are very similar to those of the commutative 
theory and do not change the micro-causal structure. The non-planar diagrams 
introduce extra momentum dependent phase factors through the $\vp_\star^3$ 
term. These phase factors play an important role in modifying the properties 
of $G_2(\l_i, x)$. The relevant Feynman diagrams are shown in \fig{feyndiags}.

We will concentrate exclusively on the non-planar contribution to the 
commutator. The phase factor originating from the $\vp_\star^3$ 
interaction for the non-planar graph is 
\eqn{phase}{
P(p_1,p_2) = e^{i \, p_1^\mu \, \th_{\mu \nu} \, p_2^\nu} \ ,}
Ignoring numerical coefficients we can write
\eqn{mtwoa}{
\CM^{(2)}_{np} \sim \int_0^\infty \, d\l_1\, \int_0^\infty \, d\l_2 \, 
\[G_2^{(np)} (\l_1, \l_2,x) -G_2^{(np)} (\l_1, \l_2,-x) \] \ , }
where 
\eqn{gtwoa}{
G_2^{(np)} (\l_1, \l_2,x) =  \exp \(
i \, { x^2 \over 2u_x} \, \(\l_1 +\l_2\)  -i \, {m^2 u_x \over 2}\, 
 \( {1\over \l_1} + {1\over \l_2} \)   \) \, H_2(\l_1,\l_2, x) \ , }
 is of the general form described in \intgnd, and the kernel $H_2(\l_1,\l_2,x)$  is 
 \eqn{htwoa}{
 H_2(\l_1,\l_2,x) = \int \, {d^4{\vec p}_1 \, d^4{\vec p}_2 \over (2 \, \pi)^{10}}
 \, e^{- i \, {u_x \over 2 \, \l_1} \, \( {\vec p}_1 - \l_1 \, {{\vec x} \over u_x} \)^2 
 -  i \, {u_x \over 2 \, \l_2} \, \( {\vec p}_2 - \l_2 \, {{\vec x} \over u_x} \)^2 }
 \; P(p_1,p_2) \;  Q \ . }
 $Q$ is a function of the momenta that grows like a polynomial and hence doesn't
 play any role in our discussions. Readers interested in its precise form can 
 consult \cite{Chu:2005nb}.  In deriving the above  we have made use of the identity
\identy\ given above to convert the momentum integrals over $d^5\bp_{1,2}$ 
to integrals over  $d^4{\vec p}_{1,2}$ and over the light cone component 
labeled here as $\l_{1,2}$ respectively. Note that while we have lumped a 
lot of the details of the interaction in the factor $Q$, we are keeping the 
phase factor $P(p_1,p_2)$ explicitly, since it is an exponential function 
of the momenta. 

We now analyze the large momentum behaviour of \htwoa\  by completing 
the squares and performing the resulting integral (setting $Q = 1$).
So far our discussion has been independent of the precise nature 
of the \nc\ interaction. We will see that there is a distinction between the 
\snc\ and \lnc\ cases. The essential point is that the large momentum 
behaviour of \htwoa\ depends on the nature of the \nc\ interaction, 
and so we will treat these two cases separately.

\newsubsection{Spacelike non-commmutativity}

For \snc\ theories we will choose $\th^{\mu \nu} \neq 0$ only along the spatial 
directions transverse to our light cone coordinates, thus having $\th^{ij}\neq 0$. 
So we can write \htwoa\ as 
\eqn{htwob}{\eqalign{
H_2(\l_1,\l_2,x) & \sim  \int \, d^4{\vec p}_1 \, d^4{\vec p}_2 \, 
 e^{- i \, {u_x \over 2 \, \l_1} \, \( {\vec p}_1 - \l_1 \, {{\vec x} \over u_x} \)^2 
 -  i \, {u_x \over 2 \, \l_2} \, \( {\vec p}_2 - \l_2 \, {{\vec x} \over u_x} \)^2 + 
i \, p_1^i \,\th_{ij}\, p_2^j } \ ,  \cr 
& \sim e^{- i \, {1\over u_x} \, {\vec x}_{nc}^2  \, \( \l_1 + \l_2 \) } \ .}}
One interesting aspect is that the large momentum behaviour of 
$H_2(\l_1,\l_2,x)$ is independent of the \nc\ parameter $\th$.
Using the expressions for $G_2^{(np)}$ we find at the end of the day,
\eqn{mtwob}{
\CM^{(2)}_{np} \sim  \int_0^\infty \,  d\l_1 \, d\l_2 \, \; \(
e^{ i \, \( \l_1 + \l_2 \) \, {x_c^2 \over 2 \, u_x} } - e^{ - i \, \( \l_1 + \l_2 \) \, 
{x_c^2 \over 2 \, u_x} } \) \ . } 
In contrast to the tree-level result, we see that the convergence properties 
at one-loop (for non-planar contributions) depend only on the light cone 
defined with respect to the commuting directions in spacetime. Once again 
going through the possible contour rotations leads us to conclude that 
the commutator is required in this case to vanish outside the light-wedge 
as in \mtwo.

\newsubsection{Lightlike non-commutativity}

Here we choose the non zero components of $\theta^{\mu \nu}$ to 
be those with $\th_{v i} = \th^i \neq 0$, with $i$ denoting the spatial directions 
in $\R^4 \subset \R^{5,1}$. This in particular implies 
that the phase factor will be 
\eqn{phasel}{
P({\vec p}_1, {\vec p}_2 ) = e^{i \, {\vec \th} \cdot  \( \l_1  \,{\vec p}_2 + 
\l_2 \, {\vec p}_1 \)} \ . }
Hence calculating the kernel $H_2(\l_1,\l_2,x)$ reduces to
\eqn{lhtwob}{\eqalign{
H_2(\l_1,\l_2,x) & \sim  \int \, d^4{\vec p}_1 \, d^4{\vec p}_2 \, 
 e^{- i \, {u_x \over 2 \, \l_1} \, \( {\vec p}_1 - \l_1 \, {{\vec x} \over u_x} \)^2 
 -  i \, {u_x \over 2 \, \l_2} \, \( {\vec p}_2 - \l_2 \, {{\vec x} \over u_x} \)^2 + 
i \,  {\vec \th} .  \( \l_1  {\vec p}_2 + 
\l_2 {\vec p}_1 \) }\ ,  \cr 
& =  \int \, d^4{\vec p}_1 \,  
\exp \(- i \ {u_x \over 2 \, \l_1} \,\( {\vec p}_1 - {\l_1  \over u_x } {\vec x} - 
 {{\l_1 \, \l_2} \over u_x}\, {\vec \th} \)^2 + i \, {u_x  \over 2 \, \l_1}  \, 
\({\l_1 \over u_x} \, {\vec x}  -  {\l_1 \, \l_2 \over u_x} \, {\vec \th} \)^2 -  
i \, { \l_1 \over 2 \, u_x} \vec{x}^2 \) \, \cr 
& \;\;\; \times \, d^4{\vec p}_2 \,  \exp \(- i \ {u_x \over 2 \, \l_2} \,\( {\vec p}_2 - 
{\l_2  \over u_x } {\vec x} - 
 {{\l_1 \, \l_2} \over u_x} \,{\vec \th} \)^2 + i \, {u_x  \over 2 \, \l_2}  \, 
\({\l_2 \over u_x} \, {\vec x}  -  {\l_1 \, \l_2 \over u_x} \,{\vec \th} \)^2 -  
i \, { \l_2 \over 2 \, u_x} \vec{x}^2 \) \ , }}
The integrals over $d^4{\vec p}_{1,2}$ are easily done, but unlike the 
case of spacelike non-commutativity discussed previously we find that the 
dominant term for large $\l_i$ depends explicitly on $\th$. In fact, we have 
\eqn{gtwobl}{
G_2^{(np)}(\l_1,\l_2,x) \sim \int_0^\infty \,  d\l_1 \, d\l_2 \, \exp \( i {{\vec \th}
 \cdot {\vec \th}
\over 2 \, u_x } \, \l_1 \, \l_2 \, \(\l_1 + \l_2\)  - i \, {m^2 \, u_x \over 2} \, 
\({1\over \l_1} + {1\over \l_2} \) \) \ . }
In writing the above we have retained only the dominant contribution
at large momentum (for it is the part that determines the convergence
properties).  If we consider the contour rotations for the $\l_{1,2}$
integrals, continuing $\l_{1,2} \to i \, \l_{1,2}$ leaves us with a
divergent integral (assuming without loss of generality $u_x >0$) for
$G^{(np)}_2(\l_1,\l_2,x)$. One might wonder about continuing into the
lower half-plane \ie, $\l_{1,2} \to - i\, \l_{1,2}$; this doesn't help
because the integral then diverges at small
$\l_i$. Hence there is no contour rotation that will allow us to obtain a
finite answer, so we are led to conclude that the correct
micro-causality condition is as given in \lmicro.

\Appendix{Non-distinguishing property the spacetime}
\label{appC}

In this appendix, we demonstrate directly that the spacetime \ndgeo,
\eqn{ndgeoC}{
ds^2 =  {r^2 } \, \( -2 \, du \, dv  - \b^2 \, r^4  \, du^2+ 
dx_1^2 + dx_2^2 +  {dr^2 \over r^4} + {1\over r^2} \, d\Omega_5^2 \) \ , 
}
is non-distinguishing.  In particular, we
exhibit a causal curve connecting a point 
$P= (u_0, v_0, r_0, x_0, \Om_0)$ in the spacetime to any point 
$Q = (u_0 +\e, v_1, r_1, x_1, \Om_1)$, 
with $\e >0$ arbitrarily small, 
and arbitrary values of $(v_1,r_1,x_1, \Om_1)$,
where $x$ and $\Om$ capture the 2 transverse directions $(x_1,x_2)$ and
the 5 angular directions of the $\S^5$, respectively.

Since along any future directed causal curve the coordinate $u$ must increase,
we can parameterise our curve by $u$, so that
\eqn{ccurve}{
\g(u)=\(u,v(u),r(u),x(u),\Om(u)\)
}
Then the causality condition implies that for all $u$,
\eqn{causalitycond}{
2 \, \vd + \b^2 \, r^4 \ge \xd{}^2 + {\rd^2 \over r^4} + {\Omd^2 \over r^2}
}
where we define $\dot{} \equiv {d \over du}$.
For simplicity, we break up the curve into five components; the first and last
being null curves in the $(r,v)$ plane, while the middle three move
in the $x$, $\Om$, and $v$ directions, respectively.  Letting each component
take equal $u$ interval, these curves, $\g_i$ with $i=1,\ldots,5$, join the points
\eqn{points}{\eqalign{
P &= (u_0, v_0, r_0, x_0, \Om_0) \cr
p_1 &= (u_0+{\e \over 5}, v_2, R, x_0, \Om_0) \cr
p_2 &= (u_0+{2\, \e \over 5}, v_2, R, x_1, \Om_0) \cr
p_3 &= (u_0+{3\, \e \over 5}, v_2, R, x_1, \Om_1) \cr
p_4 &= (u_0+{4 \, \e \over 5}, v_3, R, x_1, \Om_1) \cr
Q &= (u_0 +\e, v_1, r_1, x_1, \Om_1)
}}
with the quantities $v_2, v_3$ and $R$ as defined below.
Choosing this joining to be linear, we can write the component
curves explicitly as
\eqn{ccurves}{\eqalign{
\g_1(u) &= \(u, v_a(u), r_a(u), x_0, \Om_0\) \cr
\g_2(u) &= \(u, v_2, R, 
 {x_1-x_0 \over \e/5} \, u + 
 {x_0 \(u_0+{2\e \over 5}\)-x_1\(u_0+{\e \over 5}\)\over  \e/5} , \Om_0\)  \cr
\g_3(u) &= \(u, v_2, R, x_1,
 {\Om_1-\Om_0 \over \e/5} \, u + 
 {\Om_0 \(u_0+{3\e \over 5}\)-\Om_1\(u_0+{2\e \over 5}\)\over  \e/5}\)  \cr
\g_4(u) &= \(u, {v_3-v_2 \over \e/5} \, u + 
 {v_2 \(u_0+{4\e \over 5}\)-v_3\(u_0+{3\e \over 5}\)\over  \e/5}, 
 R, x_1, \Om_1\) \cr
\g_5(u) &= \(u, v_b(u), r_b(u), x_1, \Om_1\) 
}}
with
\eqn{rv}{\eqalign{
r_a(u) &\equiv \rho_a \, u + \mu_a \cr
v_a(u) &\equiv v_0+{\rho_a \over 6} \( {1\over r_0^3} - {1 \over r_a(u)^3} \)
-{\b^2 \over 10 \rho_a} \( r_a(u)^5 - r_0^5 \) \cr
r_b(u) &\equiv \rho_b \, u + \mu_b \cr
v_b(u) &\equiv v_3+{\rho_b \over 6} \( {1\over R^3} - {1 \over r_b(u)^3} \)
-{\b^2 \over 10 \rho_b} \( r_b(u)^5 - R^5 \) \ ,
}}
where we define the constants
\eqn{consts}{\eqalign{
\rho_a &= {R-r_0 \over \e/5}\ , \qquad 
\mu_a = {r_0  \(u_0+{\e \over 5}\) - R  \, u_0 \over \e/5} \cr
\rho_b &= - {R-r_1 \over \e/5}\ , \qquad 
\mu_b = {R  \(u_0+\e \) - r_1  \, \(u_0+{4\e \over 5}\) \over \e/5} \cr
v_2 &= v_0 + {\rho_a \over 6} \( {1\over r_0^3} - {1 \over R^3} \)
-{\b^2 \over 10 \rho_a} \( R^5 - r_0^5 \)  \cr
v_3 &= v_1 + {\rho_b \over 6} \( {1\over r_1^3} - {1 \over R^3} \)
-{\b^2 \over 10 \rho_b} \( R^5 - r_1^5 \)  
}}
These ensure that the curve is connected, joining the points \points, 
and \rv\ ensure that $\g_1$ and $\g_5$ are null.
Finally, to guarantee causality of $\g_2, \g_3$ and $\g_4$, 
we must choose $R$ bounded from below by
\eqn{Rmin}{
R \ge {\rm max} \left\{ \( {x_1 - x_0 \over \b \, \e/5} \)^{\! 1/2} ,
\( {\Om_1 - \Om_0 \over \b \, \e/5} \)^{\! 1/3} ,
\( {v_2 - v_3 \over \b^2 \, \e/10} \)^{\! 1/4} \right\} \ .
}
Physically, we want to take advantage of the large-$r$ region, where the 
$-\b^2 \, r^4 \, du^2$ term in the metric \ndgeoC\ allows causality for 
large changes in the other coordinates.  A schematic representation 
of such  a causal curve in $\{u,v,r\}$ space is sketched in \fig{nondistlc}.
Note that this is only possible for nonzero $\b$.

Hence, a crucial feature of our construction of the causal curve $\g$
is that it involves 
arbitrarily large radii, scaling as inverse positive power of $\D u = \e$.
In other words, it is exactly due to the asymptotic region that the 
spacetime is non-distinguishing.  As commented above, this bears
two important consequences:  it lets us extract this causal property
directly from the dual boundary theory, and it pacifies possible 
causality-violating quantum fluctuations.

Likewise, in the tamer case of the spacetime \spncmet, 
the holographic dual to the $\CN=4$ SYM with spacelike
non-commutativity, it is important that the light wedge structure appears
only asymptotically.  For otherwise, if the spacetime exhibited 
a light-wedge structure even in the interior, then any two points separated
only in the \nc\ directions (along the wedge) would have identical past/future 
sets---the spacetime would be non-distinguishing.  But we know that this
is not the case, as \spncmet\ admits a time function: it is stably causal
and therefore distinguishing.

\Appendix{Other solutions generated by null Melvin twist}
\label{appB}

In this appendix we write down metrics generated by applying the 
null Melvin twist transformations on other geometries. 

\newsubsection{Twisting along angular isometries}

Instead of performing the null Melvin twist on the planar 
directions $(x_1,x_2)$ we could instead have used some 
angular directions. The most general such configuration can be 
generated by starting with the D3-brane solution written 
as \dtbrane, with
\eqn{dtalt}{\eqalign{
dx_1^2 + dx_2^2 &= d\rho^2 + \rho^2 \, d\phi_1^2 \cr
d\Omega_5^2 & =  d\chi^2 + \cos^2 \chi \,
d\phi_2^2  + {1\over 4} \, \sin^2 \chi \, \(d\theta^2 + d\phi_3^2
+ d\psi^2 + 2 \, \cos \theta \, d\phi_3 \, d\psi \)
}}
We can in this case perform the null Melvin twist by 
\eqn{altwist}{
d\phi_i \to \;\; d\phi_i +   \, dy  \ , \qquad {\rm for} \;\; i=1,2,3 \ .}
This leads to the metric
\eqn{dtmtalt}{\eqalign{
ds^2 &= -H^{-{3 \over 2}} \, \b^2 \, \(\rho^2 + r^2 \, H \) \,
(dt+dy)^2 + {1\over \sqrt{H}}\, \(-dt^2 + dy^2 + d\rho^2 + \rho^2 \,
d\phi_1^2\) \cr
& \qquad  + \sqrt{H} \, \( dr^2 + r^2 \[ d\chi^2 + \cos^2 \chi \,
d\phi_2^2  + {1\over 4} \, \sin^2 \chi \, \(d\theta^2 + d\phi_3^2
+ d\psi^2 + 2 \, \cos \theta \, d\phi_3 \, d\psi \) \] \)  \cr 
B &= \b \, (dt + dy) \, \wedge \,
\({\rho^2 \over H} \, d\phi_1 + r^2 \, \cos^2 \chi \, d\phi_2 +
{r^2 \over 4} \, \sin^2 \chi \, (d\phi_3 + \cos\theta \, d\psi) \) 
}}
While the metric \dtmtalt\ is interesting in its own right, we 
will mostly focus on the simpler case presented in \dtnmtw.

\newsubsection{Non-extremal null Melvin twisted D3-brane geometry}

To obtain the supergravity dual to thermal \lnc\ $\CN=4$ SYM, we 
need to start with a non-extremal D3-brane geometry and carry out 
the null Melvin twist duality transformation. Recall that the metric 
for the non-extremal D3-brane takes the form
\eqn{nedtb}{
ds^2 = {1\over \sqrt{H(r)}} \, \( - f(r) \, dt^2 + dy^2 + dx_1^2 + dx_2^2 \)
+ \sqrt{H(r)} \, \({dr^2 \over f(r)} + r^2 \, d\Omega_5^2 \) \ ,}
with
\eqn{frdef}{
f(r) = 1 - {r_0^4 \over r^4} \ . }
In this case the non-extremality of the solution implies that we no longer 
have the $SO(1,1)$ symmetry along the D3-brane world-volume. However, 
$\dya$ is still a good spacelike Killing vector and we have the requisite 
symmetries to proceed with the duality chain.

Carrying out Steps {\bf 1} through {\bf 7} outlined in Section 2.1, we
obtain the metric for the null Melvin twisted non-extremal D3-brane:
\eqn{nextdtnmt}{\eqalign{
ds^2 &= 
{H \left( -f dt^2 + dy^2 + dx_1^2 + dx_2^2 \right)
- \b^2 \, f \, (dt+dy)^2 
+ {1 \over 2}\, \b^2  \, (1-f)(dx_1 - dx_2)^2  \over \sqrt{H} \, 
\left(H + \b^2 \, (1-f) \right)} \cr
& \qquad \qquad+  \sqrt{H}\left({dr^2 \over f(r)} + r^2 \, d \Omega_5^2\right)  
}}
Here as in \dtnmtw, we have twisted along the non-compact directions
$(x_1,x_2)$ for simplicity. We have also refrained from writing down
the explicit expressions for the p-form fields supporting the
solution. This solution was also discussed in \cite{Nayak:2004rc}.

The metric is a bit more complicated than \dtnmtw, but it is easy to
check that in the decoupling limit (replacing $H(r) \to R^4/r^4$)
 the asymptotic form of the metric coincides with that of
\ndgeo. Since it is the asymptotic region that is responsible for the
non-distinguishing character of the spacetime, we conclude that
\nextdtnmt\ is also a non-distinguishing spacetime.  That is, since
the decoupling limit of 
\nextdtnmt\ and \ndgeo\ have a similar large $r$ behaviour, two
points which are sufficiently far away from the black hole ought to
have the same causal future/past, implying that the spacetime is
non-distinguishing. 

\newsubsection{Null Melvin twist of Klebanov-Strassler}

The Klebanov-Strassler solution is given by the metric
\eqn{kbsrtm}{
ds^2 = h(r)^{-1/2} \, \(-dt^2 + dy^2 + dx_1^2 + dx_2^2 \) + h(r)^{1/2} \, ds_6^2 \ , }
where $ds_6^2$ denotes the metric of the deformed conifold and $r$ is a 
radial coordinate that measures the size of the base. The solution for the 
warp factor $h(r)$ is given as an integral expression 
\eqn{warpks}{
h(r) = C \, \int_r^{\infty}\, dz \,{ z \coth z -1 \over \sinh^2 z } \, \(\sinh(2 z) 
- 2\, z\)^{{1\over 3}} \ , }
and the metric is supported by both five-form and three-form fluxes.
Asymptotically, this metric approaches the Klebanov-Tseytlin (KT) 
geometry~\cite{Klebanov:2000nc} with 
\eqn{ktmet}{
h(r) = {R^4 \over r^4} \, \log \({ r \over r_s} \) \ , }
where $R$ is the length scale set by the number of fractional branes
and $r_s$ the IR scale where we should revert to the original solution
\warpks.

For our purposes it will suffice to note that we have a flat metric on
$\R^{3,1}$ which is non-trivially warped. Thus we can carry out the null
Melvin twist along the translational isometries. Carrying out the
steps in the duality chain we will end up with a metric that is
analogous to \ndgeo. Apart from the fact that we have to replace the
$\S^5$ by the base of the deformed conifold we have no new change. In
the large $r$ limit, where we have the KT solution, the base is the
Einstein space $T^{1,1}$. The astute reader will have realized that we
have already taken the near horizon limit in writing the metric
\kbsrtm. So the spacetime dual to \lnc\ deformed $\CN =1$ cascade
theory is
\eqn{kslnc}{ 
ds^2 = - \, \b^2 \, h(r)^{- {3 \over 2}}\,
 \(dt + dy\)^2 + {1 \over \sqrt{h(r)}} \, \( 
-dt^2 + dy^2 + dx_1^2 + dx_2^2 \) +  \sqrt{h(r)}
\, ds_6^2 \ ,  }
which asymptotically looks like
\eqn{ktlnc}{ ds^2 = -{r^2 \over R^2 \, \sqrt{\log(r/r_s)} } \, \( -2
\, du \, dv + dx_1^2 + dx_2^2 - \b^2 \, {r^4 \over R^4 \, \log(r/r_s)
} \, \, du^2 + {R^4 \over r^4} \, \log(r/r_s) \, \(dr^2 + r^2 \,
d\Omega_5^2 \) \) \ . }
The only difference from the non-distinguishing spacetime \ndgeo\ is the 
logarithmic pieces. However, it is easy to check that these do not affect the 
causal structure of the spacetime in an essential manner.

\providecommand{\href}[2]{#2}\begingroup\raggedright\endgroup

\end{document}

%% file: lmacros.tex
\input umacros
\def\lbldef#1#2{\expandafter\gdef\csname #1\endcsname {#2}}
\def\eqn#1#2{\lbldef{#1}{(\ref{#1})}%
\begin{equation} #2 \label{#1} \end{equation}}
\def\eqalign#1{\vcenter{\openup1\jot
    \halign{\strut\span\TL & \span\TR\cr #1 \cr
   }}}


%% file: umacros.tex
\def\TL{\hfil$\displaystyle{##}$}
\def\TR{$\displaystyle{{}##}$\hfil}

\def\comment#1{}
\def\fixit#1{}

\def\mop#1{\mathop{\rm #1}\nolimits}

\def\coth{\mop{coth}}

\def\overleftrightarrow#1{\vbox{\ialign{##\crcr
     $\leftrightarrow$\crcr\noalign{\kern-0pt\nointerlineskip}
     $\hfil\displaystyle{#1}\hfil$\crcr}}}

\def\lsim{\mathrel{\mathstrut\smash{\ooalign{\raise2.5pt\hbox{$<$}\cr\lower2.5pt\hbox{$\sim$}}}}}
\def\gsim{\mathrel{\mathstrut\smash{\ooalign{\raise2.5pt\hbox{$>$}\cr\lower2.5pt\hbox{$\sim$}}}}}

\def\sqr#1#2{{\vcenter{\vbox{\hrule height.#2pt
         \hbox{\vrule width.#2pt height#1pt \kern#1pt
            \vrule width.#2pt}
         \hrule height.#2pt}}}}

\def\href#1#2{#2}  


%% file: paper.bbl
\begin{thebibliography}{10}

\bibitem{Aharony:1999ti}
O.~Aharony, S.~S. Gubser, J.~M. Maldacena, H.~Ooguri, and Y.~Oz, ``Large N
  field theories, string theory and gravity,'' {\em Phys. Rept.} {\bf 323}
  (2000) 183--386,
\href{http://www.arXiv.org/abs/hep-th/9905111}{{\tt hep-th/9905111}}.

\bibitem{Horowitz:1999gf}
G.~T. Horowitz and N.~Itzhaki, ``Black holes, shock waves, and causality in the
  AdS/CFT correspondence,'' {\em JHEP} {\bf 02} (1999) 010,
\href{http://www.arXiv.org/abs/hep-th/9901012}{{\tt hep-th/9901012}}.

\bibitem{Gao:2000ga}
S.~Gao and R.~M. Wald, ``Theorems on gravitational time delay and related
  issues,'' {\em Class. Quant. Grav.} {\bf 17} (2000) 4999--5008,
\href{http://www.arXiv.org/abs/gr-qc/0007021}{{\tt gr-qc/0007021}}.

\bibitem{Kabat:1999yq}
D.~Kabat and G.~Lifschytz, ``Gauge theory origins of supergravity causal
  structure,'' {\em JHEP} {\bf 05} (1999) 005,
\href{http://www.arXiv.org/abs/hep-th/9902073}{{\tt hep-th/9902073}}.

\bibitem{Gregory:2000an}
J.~P. Gregory and S.~F. Ross, ``Looking for event horizons using {UV/IR}
  relations,'' {\em Phys. Rev. D} {\bf 63} (2001) 104023,
\href{http://www.arXiv.org/abs/hep-th/0012135}{{\tt hep-th/0012135}}.

\bibitem{Chaichian:2002vw}
  M.~Chaichian, K.~Nishijima and A.~Tureanu, ``Spin-statistics and CPT theorems in noncommutative field theory,''  {\em Phys.\ Lett.} {\bf B568}, 146 (2003)
  \href{http://www.arXiv.org/abs/hep-th/0209008}
 {{\tt hep-th/0209008}}.
  
\bibitem{Alvarez-Gaume:2003mb}
L.~Alvarez-Gaume and M.~A. Vazquez-Mozo, ``General properties of noncommutative
  field theories,'' {\em Nucl. Phys.} {\bf B668} (2003) 293--321,
\href{http://www.arXiv.org/abs/hep-th/0305093}{{\tt hep-th/0305093}}.

\bibitem{Chu:2005nb}
C.-S. Chu, K.~Furuta, and T.~Inami, ``Locality, causality and noncommutative
  geometry,''
\href{http://www.arXiv.org/abs/hep-th/0502012}{{\tt hep-th/0502012}}.

\bibitem{Gauntlett:2002nw}
J.~P. Gauntlett, J.~B. Gutowski, C.~M. Hull, S.~Pakis, and H.~S. Reall, ``All
  supersymmetric solutions of minimal supergravity in five dimensions,'' {\em
  Class. Quant. Grav.} {\bf 20} (2003) 4587--4634,
\href{http://www.arXiv.org/abs/hep-th/0209114}{{\tt hep-th/0209114}}.

\bibitem{Gibbons:1999uv}
G.~W. Gibbons and C.~A.~R. Herdeiro, ``Supersymmetric rotating black holes and
  causality violation,'' {\em Class. Quant. Grav.} {\bf 16} (1999) 3619--3652,
\href{http://www.arXiv.org/abs/hep-th/9906098}{{\tt hep-th/9906098}}.

\bibitem{Boyda:2002ba}
E.~K. Boyda, S.~Ganguli, P.~Horava, and U.~Varadarajan, ``Holographic
  protection of chronology in universes of the Goedel type,'' {\em Phys. Rev.}
  {\bf D67} (2003) 106003,
\href{http://www.arXiv.org/abs/hep-th/0212087}{{\tt hep-th/0212087}}.

\bibitem{Flores:2002fx}
J.~L. Flores and M.~Sanchez, ``Causality and conjugate points in general plane
  waves,'' {\em Class. Quant. Grav.} {\bf 20} (2003) 2275--2292,
\href{http://www.arXiv.org/abs/gr-qc/0211086}{{\tt gr-qc/0211086}}.

\bibitem{Hubeny:2003sj}
V.~E. Hubeny, M.~Rangamani, and S.~F. Ross, ``Causal inheritance in plane wave
  quotients,'' {\em Phys. Rev.} {\bf D69} (2004) 024007,
\href{http://www.arXiv.org/abs/hep-th/0307257}{{\tt hep-th/0307257}}.

\bibitem{Flores:2004dr}
J.~L. Flores and M.~Sanchez, ``On the geometry of pp-wave type spacetimes,''
\href{http://www.arXiv.org/abs/gr-qc/0410006}{{\tt gr-qc/0410006}}.

\bibitem{Hashimoto:1999ut}
A.~Hashimoto and N.~Itzhaki, ``Non-commutative Yang-Mills and the AdS/CFT
  correspondence,'' {\em Phys. Lett.} {\bf B465} (1999) 142--147,
\href{http://www.arXiv.org/abs/hep-th/9907166}{{\tt hep-th/9907166}}.

\bibitem{Maldacena:1999mh}
J.~M. Maldacena and J.~G. Russo, ``Large N limit of non-commutative gauge
  theories,'' {\em JHEP} {\bf 09} (1999) 025,
\href{http://www.arXiv.org/abs/hep-th/9908134}{{\tt hep-th/9908134}}.

\bibitem{Chaichian:2004hb}
  M.~Chaichian and A.~Tureanu,
 ``Jost-Lehmann-Dyson representation and Froissart-Martin bound in quantum
  field theory on noncommutative space-time,''
  \href{http://www.arXiv.org/abs/hep-th/0403032}
  {{\tt hep-th/0403032}}.
  
\bibitem{Alishahiha:2003ru}
M.~Alishahiha and O.~J. Ganor, ``Twisted backgrounds, pp-waves and nonlocal
  field theories,'' {\em JHEP} {\bf 03} (2003) 006,
\href{http://www.arXiv.org/abs/hep-th/0301080}{{\tt hep-th/0301080}}.

\bibitem{Gimon:2003xk}
E.~G. Gimon, A.~Hashimoto, V.~E. Hubeny, O.~Lunin, and M.~Rangamani, ``Black
  strings in asymptotically plane wave geometries,'' {\em JHEP} {\bf 08} (2003)
  035,
\href{http://www.arXiv.org/abs/hep-th/0306131}{{\tt hep-th/0306131}}.

\bibitem{Alishahiha:2000pu}
M.~Alishahiha, Y.~Oz, and J.~G. Russo, ``Supergravity and light-like
  non-commutativity,'' {\em JHEP} {\bf 09} (2000) 002,
\href{http://www.arXiv.org/abs/hep-th/0007215}{{\tt hep-th/0007215}}.

\bibitem{Aharony:2000gz}
O.~Aharony, J.~Gomis, and T.~Mehen, ``On theories with light-like
  noncommutativity,'' {\em JHEP} {\bf 09} (2000) 023,
\href{http://www.arXiv.org/abs/hep-th/0006236}{{\tt hep-th/0006236}}.

\bibitem{Hubeny:2005ab}
V.~E. Hubeny, M.~Rangamani, and S.~F. Ross, ``Causally pathological
spacetimes are physically relevant,''
\href{http://www.arXiv.org/abs/gr-qc/0504013}{{\tt gr-qc/0504013}}.

\bibitem{Minwalla:1999px}
S.~Minwalla, M.~Van~Raamsdonk, and N.~Seiberg, ``Noncommutative perturbative
  dynamics,'' {\em JHEP} {\bf 02} (2000) 020,
\href{http://www.arXiv.org/abs/hep-th/9912072}{{\tt hep-th/9912072}}.

\bibitem{Matusis:2000jf}
A.~Matusis, L.~Susskind, and N.~Toumbas, ``The IR/UV connection in the
  non-commutative gauge theories,'' {\em JHEP} {\bf 12} (2000) 002,
\href{http://www.arXiv.org/abs/hep-th/0002075}{{\tt hep-th/0002075}}.

\bibitem{Ishibashi:1999hs}
N.~Ishibashi, S.~Iso, H.~Kawai, and Y.~Kitazawa, ``Wilson loops in
  noncommutative Yang-Mills,'' {\em Nucl. Phys.} {\bf B573} (2000) 573--593,
\href{http://www.arXiv.org/abs/hep-th/9910004}{{\tt hep-th/9910004}}.

\bibitem{Gross:2000ba}
D.~J. Gross, A.~Hashimoto, and N.~Itzhaki, ``Observables of non-commutative
  gauge theories,'' {\em Adv. Theor. Math. Phys.} {\bf 4} (2000) 893--928,
\href{http://www.arXiv.org/abs/hep-th/0008075}{{\tt hep-th/0008075}}.

\bibitem{Das:2000md}
S.~R. Das and S.-J. Rey, ``Open Wilson lines in noncommutative gauge theory and
  tomography of holographic dual supergravity,'' {\em Nucl. Phys.} {\bf B590}
  (2000) 453--470,
\href{http://www.arXiv.org/abs/hep-th/0008042}{{\tt hep-th/0008042}}.

\bibitem{Minwalla:1999xi}
S.~Minwalla and N.~Seiberg, ``Comments on the IIA NS5-brane,'' {\em JHEP} {\bf
  06} (1999) 007,
\href{http://www.arXiv.org/abs/hep-th/9904142}{{\tt hep-th/9904142}}.

\bibitem{Kapustin:1999ci}
A.~Kapustin, ``On the universality class of little string theories,'' {\em
  Phys. Rev.} {\bf D63} (2001) 086005,
\href{http://www.arXiv.org/abs/hep-th/9912044}{{\tt hep-th/9912044}}.


\bibitem{Chaichian:2004qk}
  M.~Chaichian, M.~N.~Mnatsakanova, K.~Nishijima, A.~Tureanu and Y.~S.~Vernov,
  ``Towards an axiomatic formulation of noncommutative quantum field  theory,''
  \href{http://www.arXiv.org/abs/hep-th/0402212}
  {{\tt arXiv:hep-th/0402212}}.
  
\bibitem{Franco:2004gx}
  D.~H.~T.~Franco and C.~M.~M.~Polito,
  ``A new derivation of the CPT and spin-statistics theorems in non-commutative
  field theories,''
  \href{http://www.arXiv.org/abs/hep-th/0403028}
  {{\tt hep-th/0403028}}.
  
\bibitem{Franco:2004fp}
  D.~H.~T.~Franco,
 ``On the Borchers class of a non-commutative field,''
  J.\ Phys.\ A {\bf 38}, 5799 (2005)
  \href{http://www.arXiv.org/abs/hep-th/0404029}
  {{\tt hep-th/0404029}}.
  
  \bibitem{Hashimoto:2004pb}
  A.~Hashimoto and K.~Thomas,
  ``Dualities, twists, and gauge theories with non-constant
  non-commutativity,''
  JHEP {\bf 0501}, 033 (2005)
  \href{http://www.arXiv.org/abs/hep-th/0410123}
  {{\tt hep-th/0410123}}.
  
\bibitem{Maldacena:1997re}
J.~M. Maldacena, ``The large N limit of superconformal field theories and
  supergravity,'' {\em Adv. Theor. Math. Phys.} {\bf 2} (1998) 231--252,
\href{http://www.arXiv.org/abs/hep-th/9711200}{{\tt hep-th/9711200}}.

\bibitem{Seiberg:1999vs}
N.~Seiberg and E.~Witten, ``String theory and noncommutative geometry,'' {\em
  JHEP} {\bf 09} (1999) 032,
\href{http://www.arXiv.org/abs/hep-th/9908142}{{\tt hep-th/9908142}}.

\bibitem{Hubeny:2002zr}
V.~E. Hubeny and M.~Rangamani, ``Causal structures of pp-waves,'' {\em JHEP}
  {\bf 12} (2002) 043,
\href{http://www.arXiv.org/abs/hep-th/0211195}{{\tt hep-th/0211195}}.

\bibitem{Seiberg:2000ms}
N.~Seiberg, L.~Susskind, and N.~Toumbas, ``Strings in background electric
  field, space/time noncommutativity and a new noncritical string theory,''
  {\em JHEP} {\bf 06} (2000) 021,
\href{http://www.arXiv.org/abs/hep-th/0005040}{{\tt hep-th/0005040}}.

\bibitem{Susskind:1998dq}
L.~Susskind and E.~Witten, ``The holographic bound in anti-de Sitter space,''
\href{http://www.arXiv.org/abs/hep-th/9805114}{{\tt hep-th/9805114}}.

\bibitem{Klebanov:2000hb}
I.~R. Klebanov and M.~J. Strassler, ``Supergravity and a confining gauge
  theory: Duality cascades and $\chi$SB-resolution of naked singularities,''
  {\em JHEP} {\bf 08} (2000) 052,
\href{http://www.arXiv.org/abs/hep-th/0007191}{{\tt hep-th/0007191}}.

\bibitem{Rozali:2000np}
M.~Rozali and M.~Van~Raamsdonk, ``Gauge invariant correlators in
  non-commutative gauge theory,'' {\em Nucl. Phys.} {\bf B608} (2001) 103--124,
\href{http://www.arXiv.org/abs/hep-th/0012065}{{\tt hep-th/0012065}}.

\bibitem{Nayak:2004rc}
R.~R. Nayak, K.~L. Panigrahi, and S.~Siwach, ``Brane solutions with / without
  rotation in pp-wave spacetime,'' {\em Nucl. Phys.} {\bf B698} (2004)
  149--162,
\href{http://www.arXiv.org/abs/hep-th/0405124}{{\tt hep-th/0405124}}.

\bibitem{Klebanov:2000nc}
I.~R. Klebanov and A.~A. Tseytlin, ``Gravity duals of supersymmetric SU(N) x
  SU(N+M) gauge theories,'' {\em Nucl. Phys.} {\bf B578} (2000) 123--138,
\href{http://www.arXiv.org/abs/hep-th/0002159}{{\tt hep-th/0002159}}.

\end{thebibliography}
